\documentclass[10pt,onecolumn,aps,prd,preprintnumbers,showpacs,superscriptaddress,nofootinbib,amsmath,amssymb,floats,floatfix,showkeys,notitlepage,longbibliography]{revtex4-2}

\usepackage{xcolor}
\usepackage[commandnameprefix=always]{changes}
\usepackage{orcidlink}
\usepackage{comment}
\usepackage{lipsum}
\usepackage{graphicx}
\usepackage{subfigure}
\usepackage{palatino}
\usepackage{graphicx,subfigure,epsfig}
\usepackage{hyperref}
\usepackage{multirow}
\usepackage{booktabs}
\usepackage[normalem]{ulem}
\useunder{\uline}{\ul}{}

\hypersetup{
    colorlinks=true,
    citecolor=blue,
    linkcolor=red,
    filecolor=magenta,
    urlcolor=cyan,}

\begin{document}

\title{Quasinormal modes of Kerr-like black bounce spacetime}
\author{Yi Yang}
\email{yiyang@mail.gufe.edu.cn}
\affiliation{School of Mathematics and Statistics, \\
Guizhou University of Finance and Economics, Guiyang, 550025, China}

\author{Dong Liu}
\email{dongliuvv@yeah.net}
\affiliation{Department of Physics, Guizhou Minzu University, Guiyang, 550025, China}

\author{Ali \"Ovg\"un}
\email{ali.ovgun@emu.edu.tr}
\affiliation{
Physics Department, Eastern Mediterranean University, Famagusta, 99628 North Cyprus via Mersin 10, Turkey.}

\author{Zheng-Wen Long}
\email{zwlong@gzu.edu.cn}
\affiliation{College of Physics, Guizhou University, Guiyang, 550025, China}

\author{Zhaoyi Xu}
\email{zyxu@gzu.edu.cn}
\affiliation{College of Physics, Guizhou University, Guiyang, 550025, China}


\begin{abstract}
We investigate the quasinormal mode (QNM) spectrum of a Kerr-like black-bounce spacetime under massive scalar-field perturbations. Starting from the Kerr-like deformation of the Simpson--Visser black-bounce geometry, we derive the corresponding radial and angular equations and obtain the effective potential governing scalar perturbations. The QNM frequencies are computed by means of the P\"oschl--Teller potential approximation and the semi-analytic WKB method (up to sixth order), and we demonstrate reasonable agreement between these two approaches. We then analyze in detail how the QNM spectrum depends on the spin parameter $a$, the bounce parameter $p$ that interpolates between black-hole and wormhole geometries, and the scalar-field mass $\mu$. Our results show that increasing the spin parameter $a$ raises the oscillation frequency, while increasing the bounce parameter $p$ lowers it, and in both cases the damping rate decreases. Moreover, the mass of the scalar field has a non-negligible impact on the ringdown spectrum. These features suggest that rotating black-bounce geometries may leave distinct imprints in the ringdown phase of gravitational-wave signals, and motivate future studies of echoes and parameter estimation in the context of present and upcoming detectors.
\end{abstract}


\maketitle



\section{Introduction}
\label{sec:intro}

The first direct detections of gravitational waves (GWs) by the LIGO and Virgo Collaborations \cite{LIGOScientific:2016aoc}, together with the imaging of black-hole shadows by the Event Horizon Telescope (EHT) \cite{EventHorizonTelescope:2019dse,Akiyama:2022tyh}, have opened a new era in strong-gravity physics. For the first time, highly dynamical, strong-field regimes of gravity --- such as the merger and ringdown of compact binaries --- can be probed observationally. This allows us to test general relativity (GR) and constrain alternative theories of gravity in regimes that were completely inaccessible only a decade ago \cite{Barack:2018yly,Martinelli:2021hir,Chen:2021cts,Berti:2015itd,Cimdiker:2021cpz,Cardoso:2016rao,Barausse:2020rsu}.

Within this new observational window, the ringdown phase of a black-hole merger plays a particularly central role. After the highly non-linear merger, the remnant object relaxes to a stationary configuration by emitting damped oscillations --- the quasinormal modes (QNMs). These modes are characterized by a discrete set of complex frequencies, whose real parts represent the oscillation frequencies while their imaginary parts encode the damping rates. In GR, the QNM spectrum of an astrophysical black hole is uniquely determined by its mass and spin, as dictated by the no-hair theorems. Measuring multiple QNMs thus provides a powerful test of the Kerr nature of the remnant and of the underlying theory of gravity \cite{Qian:2021aju,Kuang:2017cgt,Fernando:2016ftj,Fernando:2012yw,Fernando:2003ai,Fernando:2009tv,Fernando:2008hb,Fernando:2014gda,Cardoso:2016oxy,Cardoso:2017cqb,Maggio:2019zyv,2022zym,Konoplya:2018yrp,Bronnikov:2019sbx,Churilova:2021tgn}.

At the same time, numerous extensions of GR and exotic compact-object models have been proposed in recent years. Alternative and modified gravity theories introduce additional degrees of freedom or new couplings, leading to field equations with different stationary solutions and distinct strong-field phenomenology \cite{Simpson:2021zfl,Simpson:2021dyo,Shankaranarayanan:2022wbx,Odintsov:2022cbm,Baker:2019gxo,Ferreira:2019xrr,Pantig:2022toh,Pantig2022,Kubiznak:2012wp,Gunasekaran:2012dq,Kerner:2008qv,Akhmedov:2006pg,Singleton:2011vh}. Among the rich landscape of non-Kerr geometries, regular black holes and wormhole-like spacetimes are especially intriguing, as they resolve curvature singularities or interpolate between black holes and wormholes, while still providing observational signatures close to those of Kerr in appropriate limits. The QNM spectrum of such objects generically deviates from the Kerr case, offering a potential avenue for tests of the no-hair conjecture and of the underlying gravity theory \cite{Daghigh:2008jz,Daghigh:2011ty,Daghigh:2020mog,Zhidenko:2003wq,Zhidenko:2005mv,Konoplya:2011qq,Chabab:2017knz,Lepe:2004kv,Gonzalez:2017shu,Okyay:2021nnh,Gonzalez:2021vwp,Panotopoulos:2020mii,Panotopoulos:2019gtn,Panotopoulos:2019qjk,Rincon:2018sgd,Ovgun:2018gwt,Berti:2009kk,Cardoso:2008bp,Andersson:2003fh,Andersson:1996cm,Andersson:1996xw,Andersson:1992scr,Andersson:1995zk,Wang:2001tk}.

A particularly elegant example of a regular spacetime is the black-bounce geometry introduced by Simpson and Visser \cite{Simpson:2018tsi}. By introducing a single parameter $a$ into the Schwarzschild solution, they constructed a geometry that interpolates smoothly between a Schwarzschild black hole ($a=0$ and $m\neq 0$) and a traversable wormhole ($a \geq 2m$). This idea has since been generalized in several directions, including new versions of the Schwarzschild-like black-bounce spacetime \cite{Lobo:2020ffi}. A broad range of physical properties of these spacetimes has been investigated: QNMs and possible gravitational-wave echoes \cite{Churilova:2019cyt,Bronnikov:2019sbx,Bronnikov:2021liv}, gravitational lensing in the weak and strong field regimes \cite{Nascimento:2020ime,Tsukamoto:2020bjm,Ovgun:2020yuv}, absorption cross sections for scalar waves \cite{Lima:2020auu}, and thin-shell wormhole constructions \cite{Lobo:2020kxn}. 
More recently, the phenomenology of Simpson--Visser and black-bounce spacetimes has been further explored from several complementary perspectives. 
For example, photon spheres, shadows, QNMs, and greybody bounds of the non-rotating Simpson--Visser black hole were analyzed in Ref.~\cite{Jha:2023wzo}. 
Thermodynamic properties, geodesic motion, shadows, and related observational signatures have also been studied for screened Simpson--Visser black holes, Simpson--Visser black holes with phantom global monopoles, charged Simpson--Visser-AdS black holes, and Simpson--Visser-AdS black holes in merger scenarios~\cite{Ahmed:2026ywu,Al-Badawi:2026dof,Ahmed:2026bwm,Kumar:2025nio}.
Rotating generalizations have also been studied: Mazza et al.\ constructed a rotating black-bounce model \cite{Mazza:2021rgq}, whose QNMs were subsequently analyzed in \cite{Franzin:2022iai}, and charged black-bounce solutions were obtained in \cite{Franzin:2021vnj}. Using a Newman--Janis-type algorithm, Xu and Tang \cite{Xu:2021lff} constructed a Kerr-like black-bounce geometry, while Guerrero et al.\ explored its lensing and shadow properties \cite{Guerrero:2021ues}. Time evolution of perturbations and retrolensing signatures in related spacetimes have also been discussed in \cite{Ou:2021efv,Tsukamoto:2022vkt}.

In this context, Kerr-like black-bounce spacetimes provide a concrete and controlled framework in which to study continuous deviations from Kerr, including configurations that are regular or wormhole-like in the interior while remaining Kerr-like at large radii. 
From an observational point of view, the ringdown signal of a perturbed rotating black-bounce geometry may carry imprints of the underlying regular core or bounce structure through shifts in the QNM spectrum. 
The scattering properties of scalar waves, such as greybody factors, can also provide complementary information about the black-bounce deformation.

The main objective of this work is to characterize the QNMs of a Kerr-like black-bounce spacetime under massive scalar-field perturbations, and to quantify how deviations from the Kerr geometry, encoded in the bounce parameter $p$ and in the choice of black-bounce profile, affect the ringdown spectrum. 
In particular, we aim to derive the radial equation and the corresponding effective potential, compute the QNM frequencies using both the P\"oschl--Teller potential approximation and the semi-analytic WKB method, compare their accuracy, and assess how the parameters $(a,p,\mu)$ shift the real and imaginary parts of the QNM spectrum.

In Sec.~\ref{bridf_review} we review the Kerr-like black-bounce spacetime, present the relevant metric functions, and derive the determinant and inverse metric components. In Sec.~\ref{scalar perturbation} we study massive scalar perturbations, derive the separated radial and angular equations, and obtain the effective potential. In Sec.~\ref{QNM}, we introduce the P\"oschl--Teller potential approximation and the WKB method, compute the QNM spectrum, and compare the two approaches. 
In Sec.~\ref{diss} we discuss the physical interpretation of the QNM spectrum and compare our results with previous studies. 
In Sec.~\ref{GBFs} we study the greybody factors of scalar waves in the same background. 
Finally, in Sec.~\ref{sec:summary} we summarize our main results and outline possible directions for future work. Throughout the paper we adopt geometrized units with $G=c=1$.

\section{Kerr-like black-bounce space-time}\label{bridf_review}
Regular spacetime metric is an interesting topic in current GR and black hole physics. It is a variant of Schwarzschild metric. In our research, we focus on black-bounce space-time which is a regular space-time that connects Schwarzschild black hole and wormhole. The Schwarzschild-like black-bounce space-time is generally read as \cite{Lobo:2020ffi}
\begin{equation}
d s^{2}=-f(r) d t^{2}+\frac{1}{g(r)} d r^{2}+H(r)\left(d \theta^{2}+\sin ^{2} \theta d \phi^{2}\right),
\end{equation}
where
\begin{equation}
\begin{aligned}
&f(r)=g(r)=1-\frac{2 \mathcal{Q}(r)}{\sqrt{H(r)}}, \\
&\mathcal{Q}(r)=\frac{M \sqrt{H(r)} r^{k}}{\left(r^{2 n}+p^{2 n}\right)^{\frac{k+1}{2 n}}}, \\
&H(r)=r^{2}+p^{2},
\end{aligned}
\end{equation}
where the non-negative parameter $p$ can determine whether space-time is a black hole or a wormhole, $M$ is the mass of compact object, $n$ and $k$ are natural number, that is, their values are 0, 1, 2, 3 $\cdots$.
When $n=1$ and $k=0$, the black-bounce space-time degenerates to Simpson-Visser model \cite{Simpson:2018tsi}. Churilova et al. studied the quasinormal modes of this  space-time, and found the echoes in the late stage \cite{Churilova:2019cyt}. In addition, we have studied another special case of Schwarzschild-like black-bounce space-time, and we also found the echoes in the late stage of quasinormal modes \cite{Yang:2021cvh}. On the other hand, we obtained Kerr-like black-bounce space-time by using Newman-Janis algorithm \cite{Xu:2021lff}

\begin{widetext}
\begin{equation}
\begin{aligned}
 ds^{2}=&-\left(1-\frac{2 M\left(r^{2}+p^{2}\right) r^{k}}{\left(r^{2 n}+p^{2 n}\right)^{\frac{k+1}{2 n}} \Sigma}\right) d t^{2}+\frac{\Sigma}{\Delta} d r^{2}
-\frac{4 a \sin ^{2} \theta}{\Sigma} \frac{M\left(r^{2}+p^{2}\right) r^{k}}{\left(r^{2 n}+p^{2 n}\right)^{\frac{k+1}{2 n}}} d td \varphi+\Sigma d \theta^{2} \\
&+\frac{\sin ^{2} \theta}{\Sigma}\left(\left(r^{2}+p^{2}+a^{2}\right)^{2}-a^{2} \Delta \sin ^{2} \theta\right) d \varphi^{2},
\end{aligned}
\end{equation}
where
\begin{equation}
\Delta=r^{2}+a^{2}+p^2-\frac{2 M r^{k}\left(r^{2}+p^{2}\right)}{\left(r^{2 n}+p^{2 n}\right)^{\frac{k+1}{2 n}}}, \quad \Sigma=r^{2}+p^2+a^{2} \cos ^{2}\theta.
\end{equation}
\end{widetext}
For $n=1,k=0,a=0,$ the rotating black bounce spacetime degenerates to Schwarzschild-like black bounce space-time \cite{Simpson:2018tsi}. If the parameters $n = 1, k = 0, p = 0,$ we can get kerr black hole metric. In Ref. \cite{Mazza:2021rgq}, Mazza et al. studied a special case of rotating black bounce space-time ($n=1,k=0$), and they further studied the quasinormal modes of this space-time \cite{Franzin:2022iai}.

In this work, we study the quasinormal modes of rotating black bounce spacetime for $n=2,k=0$. Therefore, the rotating black bounce space-time metric can be written as
\begin{widetext}
\begin{equation}\label{line}
d s^{2}=-\left(1-\frac{2 M (r^2+p^2)}{(r^4+p^4)^{1/4}\Sigma}\right) d t^{2}-\frac{4 M a (r^2+p^2) \sin ^{2}\theta}{(r^4+p^4)^{1/4}\Sigma} d t d \varphi+\frac{\Sigma}{\Delta} d r^{2}+\Sigma d \theta^{2}+\frac{A \sin ^{2}\theta}{\Sigma} d \varphi^{2},
\end{equation}
\end{widetext}
where
\begin{equation}
\begin{aligned}
&\Delta=r^{2}+a^{2}+p^2-\frac{2 M (r^2+p^2)}{(r^4+p^4)^{1/4}}, \\ &\Sigma=r^{2}+p^2+a^{2} \cos ^{2}\theta,\\
&A=\left[r^{2}+a^{2}+p^2\right]^{2}-\Delta a^{2} \sin ^{2}\theta.
\end{aligned}
\end{equation}
When the black hole spin $a = 0$, the rotating black bounce spacetime reduces to the black bounce spacetime studied in Ref. \cite{Yang:2021cvh}.
Moreover, this rotating black bounce space-time contains black-hole and wormhole branches.

In the present work, we focus on the branch $(n,k)=(2,0)$ as a simple and representative Kerr-like black-bounce profile. 
For this branch, the generalized mass function becomes
\begin{equation}
m_{2,0}(r)=\frac{M(r^2+p^2)}{(r^4+p^4)^{1/4}} .
\end{equation}
The bounce parameter $p$ then enters the radial structure in a nontrivial but analytically tractable way, and it directly affects the horizon structure and the effective potential governing scalar perturbations. 
This makes the $(n,k)=(2,0)$ branch suitable for studying how a representative black-bounce deformation modifies the QNM spectrum and the greybody factor. 
Other integer choices, such as $(n,k)=(1,1)$, also possess the Kerr limit when $p\rightarrow0$, but they lead to different radial mass profiles, extremal limits, and perturbation potentials. 
A systematic comparison among different $(n,k)$ branches would therefore require a separate analysis and will be left for future work.

According to Kerr-like black bounce space-time metric (\ref{line}), we can write its metric tensor
\begin{equation}
g_{\mu \nu}=\left(\begin{array}{cccc}
-\left(1-\frac{2 M (r^2+p^2)}{(r^4+p^4)^{1/4}\Sigma}\right)  & 0 & 0 & -\frac{2 M a (r^2+p^2) \sin ^{2}\theta}{(r^4+p^4)^{1/4}\Sigma} \\
0 & \frac{\Sigma}{\Delta} & 0 & 0 \\
0 & 0 & \Sigma & 0 \\
-\frac{2 M a (r^2+p^2) \sin ^{2}\theta}{(r^4+p^4)^{1/4}\Sigma} & 0 & 0 & \frac{A \sin ^{2} \theta}{\Sigma}
\end{array}\right),
\end{equation}
from which we can obtain the determinant of metric tensor
\begin{equation}\label{det}
g \equiv \operatorname{det}\left(g_{\mu \nu}\right)=-\Sigma^{2} \sin^{2} \theta.
\end{equation}
Therefore, the contravariant form of metric tensor $g_{\mu \nu}$ can be written as
\begin{equation}\label{contra}
g^{\mu \nu}=\left(\begin{array}{cccc}
-\frac{A}{\Sigma \Delta} & 0 & 0 & -\frac{2 M a (r^2+p^2)}{(r^4+p^4)^{1/4}\Sigma \Delta } \\
0 & \frac{\Delta}{\Sigma} & 0 & 0 \\
0 & 0 & \frac{1}{\Sigma} & 0 \\
-\frac{2 M a (r^2+p^2)}{(r^4+p^4)^{1/4}\Sigma \Delta } & 0 & 0 & \frac{\Delta-a^2\sin^2\theta}{\Sigma \Delta\sin ^{2} \theta}
\end{array}\right).
\end{equation}

\section{Scalar perturbation of the Kerr-like black-bounce space-time}\label{scalar perturbation}
In this section, we will study the scalar perturbation of Kerr-like black-bounce space-time, and derive the effective potential of scalar particles in Kerr-like black-bounce space-time. To achieve this goal, we study Klein-Gordon equation. In curved spacetime, the Klein-Gordon equation can be written as
\begin{equation}\label{kg}
\frac{1}{\sqrt{-g}} \partial \mu\left(\sqrt{-g} g^{\mu \nu} \partial \nu \Psi\right)=\mu^{2} \Psi,
\end{equation}
with $\mu$ being the mass of the scalar particle.
By bringing determinant of metric tensor (\ref{det}) and the contravariant form of metric tensor $g_{\mu \nu}$ (\ref{contra}) into Klein-Gordon equation (\ref{kg}), we can write Eq. (\ref{kg}) as

\begin{widetext}

\begin{equation}\label{eq11}
\begin{aligned}
-\frac{A}{\Sigma \Delta} \partial_{t}^{2} \Psi-\frac{2 M a (r^2+p^2)}{(r^4+p^4)^{1/4}\Sigma \Delta } \partial_{t} \partial_{\phi} \Psi&+\frac{1}{\Sigma \sin \theta} \partial_{\theta}\left(\sin \theta \partial_{\theta}\right) \Psi+ \\
&\frac{1}{\Sigma} \partial_{r}\left(\Delta \partial_{r}\right)-\frac{2 M a (r^2+p^2)}{(r^4+p^4)^{1/4}\Sigma \Delta }\partial_{\phi} \partial_{t} \Psi+\frac{\Delta-a^2\sin ^{2} \theta}{\Delta\Sigma \sin ^{2} \theta} \partial_{\phi}^{2} \Psi=\mu^{2} \Psi.
\end{aligned}
\end{equation}
\end{widetext}
To carry out separation of variables, we can adopt the following
ansatz
\begin{equation}
\Psi(r, t)=R_{l m}(r) S(\theta) e^{i m \phi} e^{-i \omega t},
\end{equation}
with $m$ being the azimuthal quantum number, and $\omega$ being the energy of the particles.
Substituting it into Eq. (\ref{eq11}), we can get the following second order partial differential radial equation
\begin{widetext}
\begin{equation}
\frac{d}{d r}\left(\Delta \frac{d R_{l m}(r)}{d r}\right)+\left[\frac{\omega^{2}\left(r^{2}+a^{2}+p^2\right)^{2}+m^{2} a^{2}}{\Delta}-\left(\omega^{2} a^{2}+\mu^{2} (r^{2}+p^2)\right)-\frac{4 M am\omega (r^2+p^2)}{(r^4+p^4)^{1/4} \Delta}\right]R_{l m}(r)=0,
\end{equation}

and the angular equation can be read as
\begin{equation}
\frac{1}{\sin \theta} \frac{d}{d \theta}\left(\sin \theta \frac{d S_{l m}}{d \theta}\right)+\left[a^{2}\left(\omega^{2}-\mu^{2}\right) \cos ^{2} \theta-\frac{m^{2}}{\sin ^{2} \theta}\right] S_{l m}(\theta)=0.
\end{equation}
\end{widetext}
The radial equation and the angular equation usually have the same eigenvalue $\Lambda_{lm}$, but the sign is opposite. These two equations can be solved by the separation constant method, so we consider the separation constant as eigenvalue $\Lambda_{lm}$. Then the second order partial differential radial equation becomes

\begin{widetext}
\begin{equation}\label{eq15}
\begin{aligned}
\Delta \frac{d}{d r}\left(\Delta \frac{d R_{l m}(r)}{d r}\right)&+\left\{m^{2} a^{2}-\frac{4 M am\omega (r^2+p^2)}{(r^4+p^4)^{1/4} }\right. \\
&\left.+ \omega^{2}\left(r^{2}+ a^{2}+p^2\right)^{2}
-\left[\mu^{2} (r^{2}+p^2)+\omega^{2} a^{2}+\Lambda_{l m}\right] \Delta\right\} R_{l m}(r)=0 .
\end{aligned}
\end{equation}
\end{widetext}
The radial solution is usually related to the free oscillation mode of the propagating field,  which has a specific frequency and its imaginary part is negative. In order to further derive the effective potential, we use the following transformation
\begin{equation}
R_{l m}(r)=\frac{\psi(r)}{\sqrt{r^{2}+a^{2}+p^{2}}},
\end{equation}
and tortoise coordinate
\begin{equation}
d r_{*}=\frac{r^{2}+a^{2}+p^{2}}{\Delta}dr.
\end{equation}
Therefore, we obtain the following second-order partial differential equation about tortoise coordinates
\begin{equation}\label{eq18}
\frac{d^{2} \psi}{d r_{*}^{2}}+\left(\omega^{2}-V_{e f f}\right) \psi=0,
\end{equation}
where $V_{e f f}$ is the effective potential, which can be written as
\begin{widetext}
\begin{equation}
V_{eff}=\frac{\Delta}{\left(a^{2}+r^{2}+p^2\right)^{2}}\left[\frac{r \Delta^{\prime}+\Delta}{a^{2}+r^{2}+p^2}-\frac{3 r^{2} \Delta}{\left(a^{2}+r^{2}+p^2\right)^{2}}-\frac{m^{2} a^{2}}{\Delta}+\frac{4 M am\omega(r^2+p^2)}{(r^4+p^4)^{1/4} \Delta}+\mu^{2} (r^{2}+p^2)+\omega^{2} a^{2}+\Lambda_{l m}\right],
\end{equation}
\end{widetext}
where $\Delta^{\prime}$ represents the derivative $\frac{d\Delta}{dr}$. 

\section{QNM of the Kerr-like black bounce space-time}\label{QNM}
In this section, we briefly introduce P\"{o}schl-Teller potential approximation and semi-analytic WKB method for calculating quasinormal modes of the Kerr-like black bounce space-time. We first introduce the relatively simple P\"{o}schl-Teller potential approximation, which was suggested by B. Mashhoon \cite{BLOME1984231,Ferrari:1984zz}. This method is mainly realized by approximating the effective potential, i.e. P\"{o}schl-Teller potential, and its usual form is
\begin{equation}
V_{P T}\left(r_{*}\right)=\frac{V_{0}}{\cosh ^{2}\left(\frac{r_{*}}{b}\right)},
\end{equation}
where
\begin{equation}
b=\frac{1}{\sqrt{-\frac{1}{2 V_{0}} \frac{d^{2} V\left(r_{0*}\right)}{d r_*^{ 2}}}},
\end{equation}
$r_{0*}$ represents the position of the maximum value $V\left(r_{*}\right)$. We use $V_0$ to denote the maximum value of the effective potential. The basic computational idea is that map Eq. (\ref{eq18}) to a Schr\"{o}dinger-like wave equation, so that the solution of equation (\ref{eq18}) satisfying the boundary condition
\begin{equation}
\begin{aligned}
&R\left(r_{*}\right) \propto e^{ i \omega r_{*}}, \text { for } r_{*} \rightarrow  \infty,\\
&R\left(r_{*}\right) \propto e^{ -i \omega r_{*}}, \text { for } r_{*} \rightarrow  -\infty.
\end{aligned}
\end{equation}
Then, Eq. (\ref{eq18}) can be mapped to the bound state of the new equation. One can analytically compute the bound states of the Schr\"{o}dinger equation, so after obtaining them, we can obtain the QNM by considering the inverse mapping. To achieve this, the following transformations need to be considered
\begin{equation}
\begin{aligned}
r_{*} & \rightarrow-i r_{*}, \\
\left(V_{0}, b\right) & \rightarrow\left(V_{0},-i b\right).
\end{aligned}
\end{equation}
To remove the complexity of the expression, we let $p=\left(V_{0}, b\right)$ and $p^{\prime}=\left(V_{0},-i b\right)$, therefore we have
\begin{equation}
V\left(-i r_{*}, p^{\prime}\right)=V\left(r_{*}, p\right).
\end{equation}
Moreover, by the definition
\begin{equation}
\begin{aligned}
&\phi\left(r_{*}, p\right)=\psi\left(-i r_{*}, p^{\prime}\right), \\
&\Omega(p) =\omega\left(p^{\prime}\right),
\end{aligned}
\end{equation}
$\phi$ satisfies the equation
\begin{equation}
\frac{d^{2} \phi}{d r_*^{2}}+\left(-\Omega^{2}+V\right) \phi=0,
\end{equation}
where
\begin{equation}
\Omega_{n}\left(V_{0}, b\right)=\frac{1}{b}\left[-\sqrt{\frac{1}{4}+V_{0} b^{2}}+\left(n+\frac{1}{2}\right)\right].
\end{equation}
Therefore, we can get the QNM
\begin{equation}
\omega_{n}\left(V_{0}, b\right)=\Omega_{n}\left(V_{0}, i b\right)=\frac{1}{b}\left[\pm \sqrt{V_{0} b^{2}-\frac{1}{4}}-\left(n+\frac{1}{2}\right) i\right]
\end{equation}
with the $n$ being the integer $n\geq0$.

On the other hand, in order to verify the accuracy of the results of the P\"{o}schl-Teller potential approximation, we also use the WKB method \cite{Konoplya:2019hlu,Konoplya:2011qq}. In the third order WKB method, we can obtain the QNM frequencies from the following equation
\begin{equation}
\omega^{2}=\left[V_{0}+\sqrt{-2 V_{0}^{''}} \Lambda(n)-i\left(n+\frac{1}{2}\right) \sqrt{-2 V_{0}^{''}}(1+\Omega(n))\right],
\end{equation}
\begin{widetext}
where
\begin{equation}
\Lambda(n)=\frac{1}{\sqrt{-2 V_{0}}}\left[\frac{1}{8}\left(\frac{V_{0}^{(4)}}{V_{0}^{\prime}}\right)\left(\frac{1}{4}+\alpha^{2}\right)-\frac{1}{288}\left(\frac{V_{0}^{\prime}}{V_{0}^{\prime}}\right)^{2}\left(7+60 \alpha^{2}\right)\right],
\end{equation}
with
\begin{equation}
\begin{aligned}
&\Omega(n)=\frac{1}{-2 V_{0}^{\prime}}\left[\frac{5}{6912}\left(\frac{V_{0}^{\prime}}{V_{0}^{\prime}}\right)^{4}\left(77+188 \alpha^{2}\right)-\frac{1}{384}\left(\frac{V_{0}^{2} V_{0}^{(4)}}{V_{0}^{3}}\right)\left(51+100 \alpha^{2}\right)+\right. \\
&\left.\frac{1}{2304}\left(\frac{V_{0}^{(4)}}{V_{0}^{\prime}}\right)^{2}\left(67+68 \alpha^{2}\right)+\frac{1}{288}\left(\frac{V_{0}^{\prime} V_{0}^{(5)}}{V_{0}^{2}}\right)\left(19+28 \alpha^{2}\right)-\frac{1}{288}\left(\frac{V_{0}^{(6)}}{V_{0}^{\prime}}\right)\left(5+4 \alpha^{2}\right)\right],
\end{aligned}
\end{equation}
\end{widetext}
where prime represents the derivation respect to $r_*$, and $\alpha=n+\frac{1}{2}$. $V_{0}^{(n)}$ denotes the $n$-order derivative of the effective potential. Moreover, sixth order WKB method is given by
\begin{equation}
\frac{i(\omega ^{2}-V_{0})}{\sqrt{-2V_{0}^{''}}}-\sum ^{6}_{i=2}\Lambda _{i}=n+\frac{1}{2}, \quad n=0,1,2, \cdots ,
\label{e32}
\end{equation}
where the WKB corrections term $\Lambda_i$ is given in Ref. \cite{Iyer:1986np}, and $V_0$ denotes the maximum of effective potential.

Tables~\ref{t1} and \ref{t2} show the comparison between the P\"oschl--Teller approximation and the WKB method for the massless and massive scalar-field perturbations, respectively. 
Here we set $M=1$, $p=1$, $l=1$, $m=n=0$. 
For both $\mu=0$ and $\mu=0.1$, the three methods give the same qualitative dependence on the spin parameter $a$. 
The real part of the frequency increases with $a$, while the damping rate $-\mathrm{Im}(\omega)$ decreases. 
Moreover, comparing Tables~I and II shows that the scalar-field mass increases the oscillation frequency and slightly weakens the damping.

To clarify the range of the spin parameter used in our numerical analysis, we further examine the horizon structure and the extremal limit of the present geometry. 
The horizon structure of the Kerr-like black-bounce spacetime is determined by the roots of $\Delta(r)=0$. 
For the present $(n,k)=(2,0)$ geometry, the extremal configuration is reached when two horizons merge, which corresponds to the double-root conditions $\Delta(r_e)=0$ and $\Delta'(r_e)=0$. 
Therefore, the extremal spin is in general a function of the bounce parameter $p$, namely $a_{\rm ext}=a_{\rm ext}(p)$, and it does not necessarily coincide with the Kerr bound $a=M$.

For $M=1$, the Kerr limit is recovered when $p=0$, where one obtains $r_e=1$ and $a_{\rm ext}=1$. 
For nonzero values of $p$, the extremal spin changes with the bounce parameter. 
For example, we find $a_{\rm ext}\simeq 1.005$ for $p=0.1$, $a_{\rm ext}\simeq 1.122$ for $p=0.5$, and $a_{\rm ext}\simeq 1.175$ for $p=1$. 
These results show that the present regular rotating geometry can admit superspinning black-hole configurations, in the sense that the extremal spin may be larger than the Kerr value $M$.

In the present work, however, we focus on the non-extremal black-hole regime. 
The numerical analysis is restricted to $a\leq 0.9$ as a conservative choice away from the near-extremal region. 
Near extremality, the effective potential becomes more sensitive to the horizon structure, and the accuracy of the WKB approximation may deteriorate. 
For this reason, the near-extremal and superspinning regimes are not explored in detail here. 
A systematic study of these regimes will be left for future work.

\begin{figure}[b!]\centering
\includegraphics[scale=0.5]{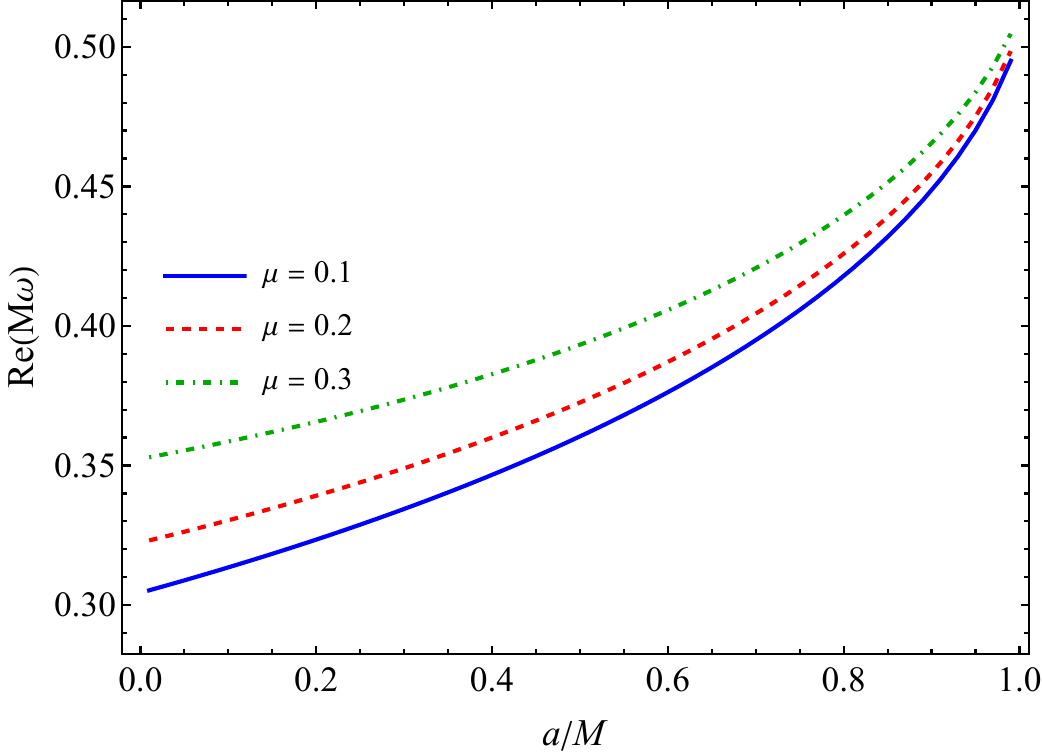}
\includegraphics[scale=0.5]{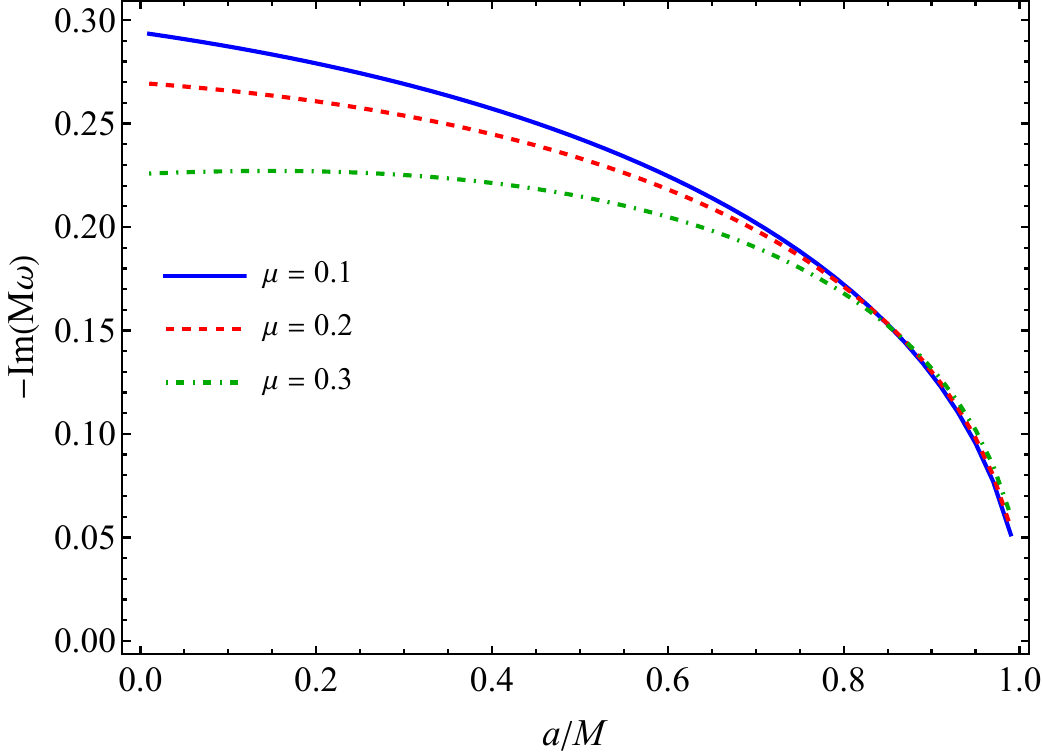}
\setlength{\abovecaptionskip}{0.1cm}
\setlength{\belowcaptionskip}{0.8cm}
\caption{QNM frequencies as functions of the spin parameter $a/M$ for different scalar-field masses $\mu$. 
The left and right panels show $\mathrm{Re}(M\omega)$ and $-\mathrm{Im}(M\omega)$, respectively. 
The parameters are fixed as $M=1$, $p=0.1$, $l=1$, $m=n=1$.}
\label{w-a}
\end{figure}

\begin{figure}[t!]\centering
\includegraphics[scale=0.5]{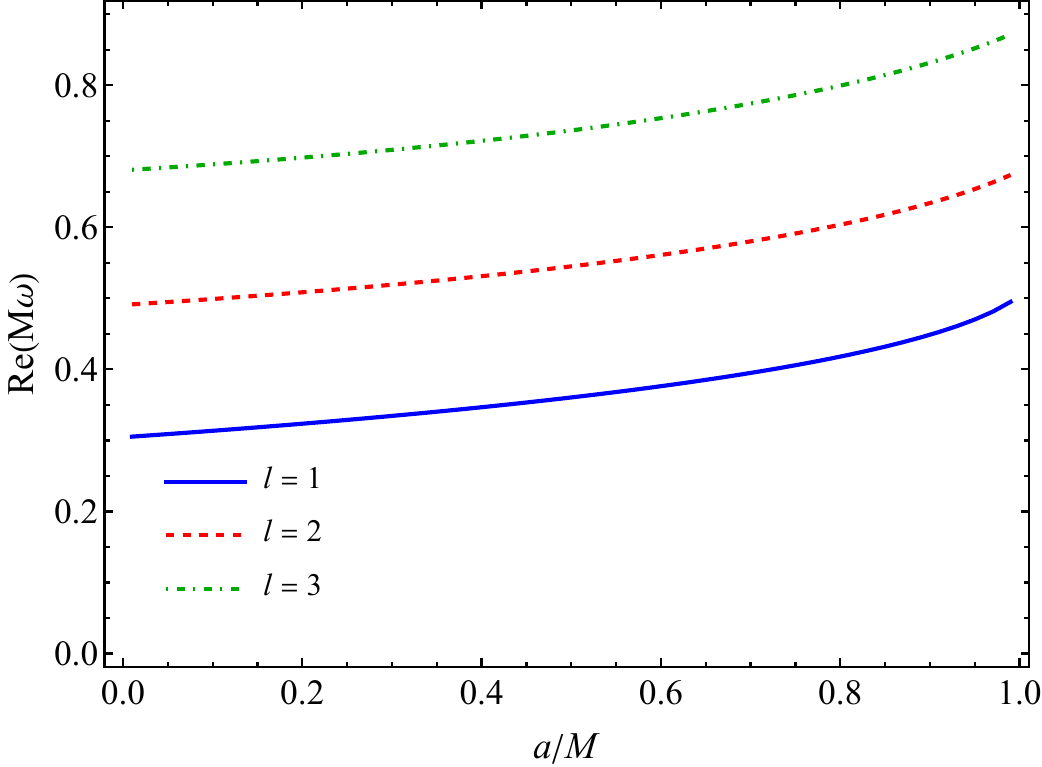}
\includegraphics[scale=0.5]{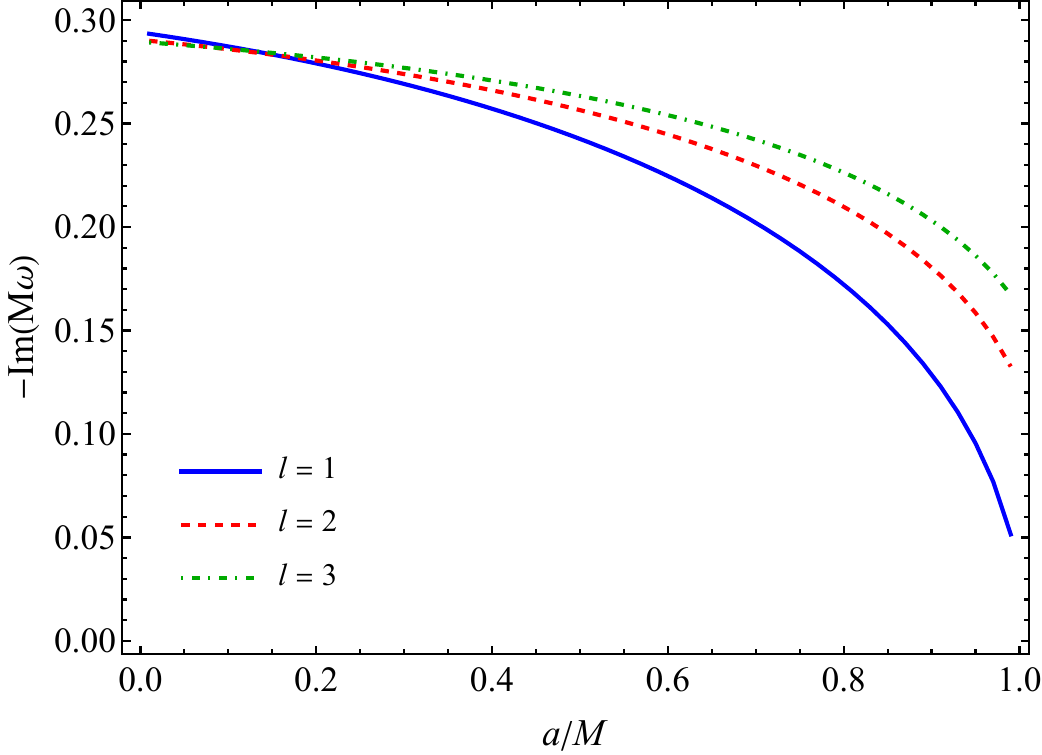}
\setlength{\abovecaptionskip}{0.1cm}
\setlength{\belowcaptionskip}{0.8cm}
\caption{
QNM frequencies as functions of the spin parameter $a/M$ for different angular numbers $l$. 
The left panel shows $\mathrm{Re}(M\omega)$, and the right panel shows $-\mathrm{Im}(M\omega)$. 
Here we set $M=1$, $p=0.1$, $\mu=0.1$, $m=n=1$.}
\label{w-l}
\end{figure}

\begin{figure}[t!]\centering
\includegraphics[scale=0.5]{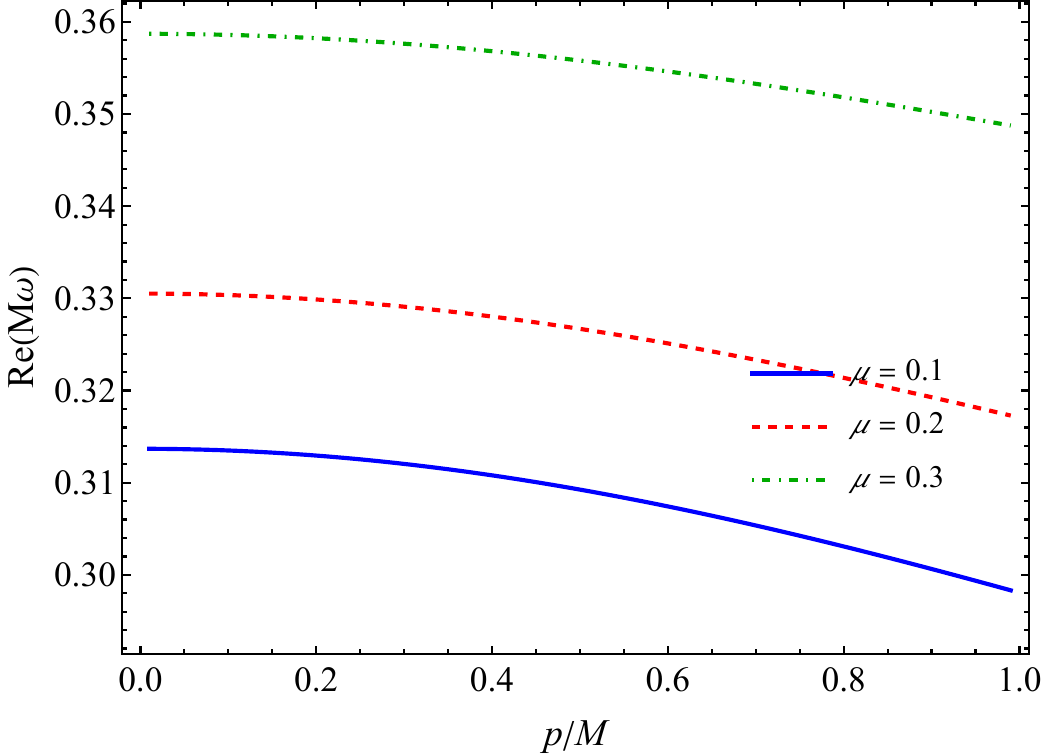}
\includegraphics[scale=0.5]{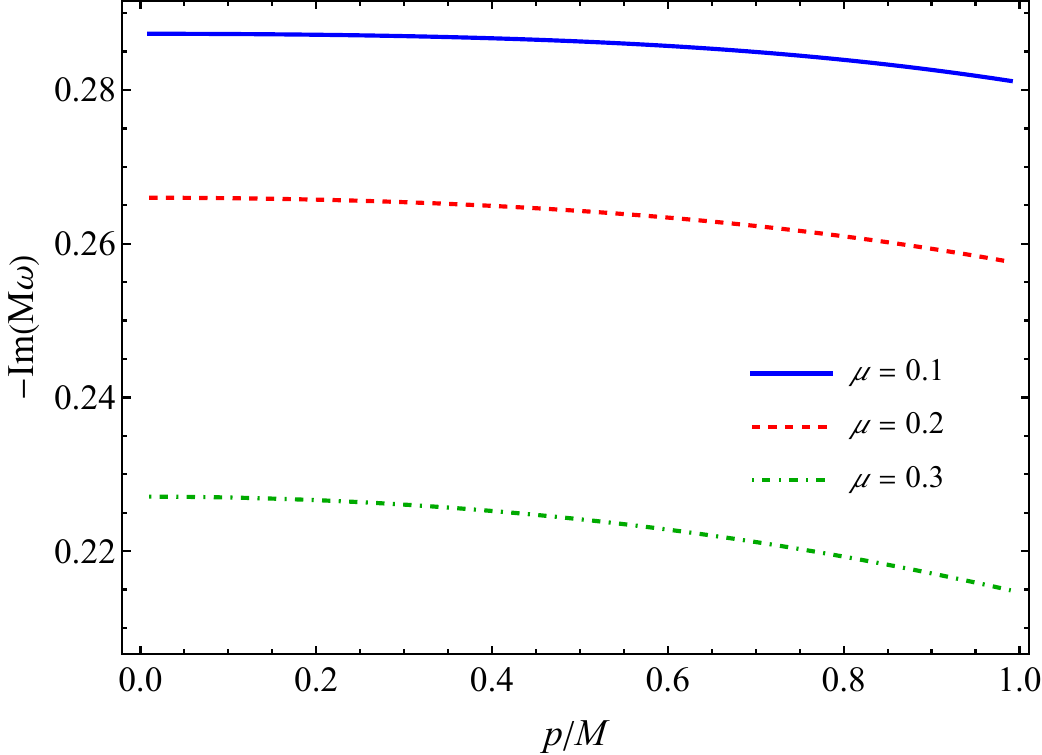}
\setlength{\abovecaptionskip}{0.1cm}
\setlength{\belowcaptionskip}{0.8cm}
\caption{
QNM frequencies as functions of the bounce parameter $p/M$ for different scalar-field masses $\mu$. 
The left panel shows $\mathrm{Re}(M\omega)$, and the right panel shows $-\mathrm{Im}(M\omega)$. 
Here we set $M=1$, $a=0.1$, $l=1$, $m=n=1$.
}
\label{w-p}
\end{figure}

Figs.~\ref{w-a} -- \ref{w-p} show the dependence of the QNM spectrum on the spin parameter, the angular number, and the bounce parameter, respectively. 
In each figure, the left panel gives the oscillation frequency $\mathrm{Re}(M\omega)$, while the right panel gives the damping rate $-\mathrm{Im}(M\omega)$.

\begin{table}[t]
\centering
\caption{
The QNM frequencies obtained by the P\"oschl--Teller approximation and the WKB method. 
Here we set $M=1$, $p=1$, $l=1$, $m=n=0$, $\mu=0$.
}
\label{tab:pt_wkb_comparison}
\begin{tabular*}{\textwidth}{@{\extracolsep{\fill}}cccc}
\toprule
$a$ & P\"oschl--Teller & 3rd-order WKB & 6th-order WKB \\
\midrule
$0.0$ & $0.283869 - 0.0983457 i$ & $0.275501 - 0.0950225 i$ & $0.278297 - 0.0947921 i$ \\
$0.1$ & $0.284020 - 0.0982771 i$ & $0.275663 - 0.0949569 i$ & $0.278464 - 0.0947201 i$ \\
$0.2$ & $0.284476 - 0.0980697 i$ & $0.276151 - 0.0947587 i$ & $0.278966 - 0.0945021 i$ \\
$0.3$ & $0.285244 - 0.0977196 i$ & $0.276972 - 0.0944229 i$ & $0.279810 - 0.0941323 i$ \\
$0.4$ & $0.286339 - 0.0972192 i$ & $0.278136 - 0.0939412 i$ & $0.281006 - 0.0936002 i$ \\
$0.5$ & $0.287781 - 0.0965574 i$ & $0.279660 - 0.0933010 i$ & $0.282566 - 0.0928912 i$ \\
$0.6$ & $0.289598 - 0.0957176 i$ & $0.281566 - 0.0924845 i$ & $0.284506 - 0.0919855 i$ \\
$0.7$ & $0.291830 - 0.0946764 i$ & $0.283883 - 0.0914677 i$ & $0.286842 - 0.0908581 i$ \\
$0.8$ & $0.294530 - 0.0934002 i$ & $0.286650 - 0.0902178 i$ & $0.289579 - 0.0894807 i$ \\
$0.9$ & $0.297771 - 0.0918398 i$ & $0.289917 - 0.0886919 i$ & $0.292706 - 0.0878257 i$ \\
\bottomrule
\end{tabular*}
\label{t1}
\end{table}

\begin{table}[t]
\centering
\caption{
The QNM frequencies obtained by the P\"oschl--Teller approximation and the WKB method for a massive scalar-field perturbation. 
Here we set $M=1$, $p=1$, $l=1$, $m=n=0$, $\mu=0.1$.
}
\label{tab:pt_wkb_mu01_comparison}
\begin{tabular*}{\textwidth}{@{\extracolsep{\fill}}cccc}
\toprule
$a$ & P\"oschl--Teller & 3rd-order WKB & 6th-order WKB \\
\midrule
$0.0$ & $0.290607 - 0.0954953 i$ & $0.280566 - 0.0920650 i$ & $0.283353 - 0.0918528 i$ \\
$0.1$ & $0.290748 - 0.0954331 i$ & $0.280722 - 0.0920064 i$ & $0.283513 - 0.0917884 i$ \\
$0.2$ & $0.291174 - 0.0952452 i$ & $0.281192 - 0.0918291 i$ & $0.283995 - 0.0915932 i$ \\
$0.3$ & $0.291893 - 0.0949275 i$ & $0.281984 - 0.0915285 i$ & $0.284805 - 0.0912617 i$ \\
$0.4$ & $0.292917 - 0.0944733 i$ & $0.283106 - 0.0910966 i$ & $0.285953 - 0.0907843 i$ \\
$0.5$ & $0.294267 - 0.0938716 i$ & $0.284575 - 0.0905214 i$ & $0.287450 - 0.0901473 i$ \\
$0.6$ & $0.295970 - 0.0931069 i$ & $0.286414 - 0.0897859 i$ & $0.289312 - 0.0893322 i$ \\
$0.7$ & $0.298062 - 0.0921568 i$ & $0.288650 - 0.0888670 i$ & $0.291552 - 0.0883163 i$ \\
$0.8$ & $0.300596 - 0.0909895 i$ & $0.291321 - 0.0877336 i$ & $0.294178 - 0.0870732 i$ \\
$0.9$ & $0.303640 - 0.0895581 i$ & $0.294477 - 0.0863444 i$ & $0.297179 - 0.0855781 i$ \\
\bottomrule
\end{tabular*}
\label{t2}
\end{table}

\section{Discussion and analysis of the QNM spectrum}\label{diss}

We now give a more detailed physical interpretation of the numerical results and compare them with previous studies of QNMs in Kerr-like, regular, and black-bounce spacetimes. 
In black-hole perturbation theory, QNM frequencies are determined by the boundary-value problem of the corresponding wave equation, and their real and imaginary parts encode the oscillation frequency and damping rate of the perturbation, respectively \cite{Berti:2009kk,Konoplya:2011qq}. 
Within the WKB picture, the QNM spectrum is mainly controlled by the local shape of the effective potential barrier. 
The real part of the frequency is primarily related to the height of the potential around its maximum, while the imaginary part is governed by the local curvature of the potential and therefore by the leakage rate of the perturbation through the barrier \cite{Schutz:1985km,Iyer:1986np,Konoplya:2019hlu}. 
This behavior is also reflected in Eq.~(28), where the QNM frequency is determined by the peak value $V_0$ and the width parameter $b$ of the effective potential. 
Thus, a higher potential barrier usually leads to a larger oscillation frequency, while a broader barrier generally gives a smaller value of $|\mathrm{Im}(\omega)|$ and hence a longer-lived mode.

Tables~I and II compare the QNM frequencies obtained by the P\"oschl--Teller approximation, the third-order WKB method, and the sixth-order WKB method. 
Table~I corresponds to the massless scalar-field perturbation with $\mu=0$, while Table~II gives a representative massive case with $\mu=0.1$. 
In both cases, we fix $M=1$, $p=1$, $l=1$, $m=n=0$. 
The three methods give the same qualitative dependence on the spin parameter $a$. 
As $a$ increases, the real part of the QNM frequency increases, whereas the damping rate $-\mathrm{Im}(\omega)$ decreases. 
Therefore, in the parameter range considered here, faster rotation enhances the oscillation frequency and makes the perturbation longer-lived. 
Moreover, comparison between Tables~I and II shows that a nonzero scalar-field mass increases $\mathrm{Re}(\omega)$ and reduces the damping rate. 
This confirms that the scalar-field mass leaves a clear imprint on the QNM spectrum and that the numerical scheme remains robust for massive scalar perturbations.

For the Kerr-like black-bounce spacetime considered here, the spin parameter $a$ affects the scalar perturbation through the radial metric function $\Delta$ and hence changes the shape of the effective potential. 
For modes with $m\neq0$, the term proportional to $am\omega$ also encodes the rotational coupling associated with frame dragging. 
In Tables~I and II, however, we focus on the axisymmetric case $m=0$, so the influence of the spin parameter mainly enters through the deformation of the background geometry. 
The increase of $\mathrm{Re}(\omega)$ with $a$ suggests that the relevant effective potential barrier is enhanced for the modes considered here. 
At the same time, the decrease of $-\mathrm{Im}(\omega)$ means that the leakage rate of the perturbation is reduced. 
As a result, the scalar perturbation oscillates faster and decays more slowly as the spin parameter increases. 
This interpretation is consistent with the general WKB expectation that the real part of the frequency is sensitive to the height of the effective potential, while the damping rate is controlled by the local curvature near the peak \cite{Iyer:1986np,Konoplya:2011qq,Konoplya:2019hlu}.

The bounce parameter $p$ has a different physical origin. 
It measures the departure from the Kerr geometry and controls the transition between black-hole-like and wormhole-like configurations. 
Increasing $p$ changes the near-core or throat region and modifies the radial structure felt by scalar perturbations. 
The decrease of $|\mathrm{Im}(\omega)|$ with increasing $p$ indicates that the damping of the scalar perturbation is weakened in the parameter range considered here. 

The role of $p$ is closely related to earlier studies of black-bounce and black-hole/wormhole transition geometries. 
For the Simpson--Visser black-bounce spacetime, it was shown that the transition between the black-hole and wormhole regimes can be accompanied by echo-like signals in the time-domain profile \cite{Churilova:2019cyt}. 
Similar echo phenomena have also been discussed in black-hole/wormhole transition models and in other black-bounce backgrounds \cite{Bronnikov:2019sbx,Bronnikov:2021liv,Yang:2021cvh}. 
However, in the present work we only compute the frequency-domain QNMs. 
A direct confirmation of echoes requires a separate time-domain analysis, which we leave for future work.

The scalar-field mass $\mu$ also plays an important role. 
The mass term modifies both the asymptotic behavior and the height of the effective potential. 
For massive perturbations, the QNM spectrum is not simply a small deformation of the massless case. 
As shown by comparing Tables~I and II, increasing $\mu$ raises the oscillation frequency and suppresses the damping rate. 
This behavior is physically natural because the mass term reduces the efficiency of wave leakage to infinity and can support longer-lived modes. 
It is also reminiscent of the quasi-resonant behavior known for massive fields in black-hole backgrounds \cite{Konoplya:2004wg}.

It is useful to compare these results with previous QNM studies of regular black holes. 
For regular black holes such as the Bardeen, Hayward, and Ay\'on-Beato--Garc\'ia black holes, earlier works showed that the regularization parameter changes the potential barrier and modifies the QNM spectrum \cite{Flachi:2012nv,Fernando:2012yw,Khoo:2025qjc,Skvortsova:2024eqi,Peng:2025htj}. 
In particular, the damping of test-field QNMs in regular black-hole spacetimes can be suppressed in comparison with the Schwarzschild case \cite{Toshmatov:2015wga}. 
Our result follows the same broad physical trend: the black-bounce parameter $p$, which regularizes the central region and controls the throat structure, tends to reduce the damping rate. 
Nevertheless, the mechanism is not identical. 
In many standard regular black-hole models, the regularization parameter modifies the effective potential and hence changes the QNM spectrum. 
In the present Kerr-like black-bounce spacetime, the bounce parameter $p$ plays a similar role by deforming the radial structure and changing the damping rate of scalar perturbations.

Recent studies have also shown that QNM spectra and greybody factors are sensitive to deformation parameters and environmental effects in different black-hole backgrounds. 
For instance, scalar perturbations, QNMs, greybody factors, and related observables have been investigated for black holes embedded in Dehnen-type dark matter halos with quintessence or string clouds~\cite{Hamil:2025pte,Al-Badawi:2024qpt}. 
Similar analyses have also been carried out for black holes in scalar-vector-tensor modified gravity, hairy black holes, and quantum-corrected Schwarzschild black holes inspired by the generalized uncertainty principle~\cite{Al-Badawi:2024iog,Al-Badawi:2024kdw,Barman:2024hwd}. 
These works further support the idea that deviations from the Kerr or Schwarzschild geometry can leave observable imprints on the effective potential, QNM spectrum, greybody factors, and shadow observables.

Compared with static black-bounce models, the present geometry contains rotation and therefore includes frame-dragging effects. 
Previous studies of the Kerr--black-bounce spacetime mainly focused on massless scalar perturbations, stability of different branches, and superradiant amplification \cite{Franzin:2022iai}. 
Here we instead focus on the massive scalar-field spectrum in the Kerr-like black-bounce geometry with $n=2$ and $k=0$, and compare the P\"oschl--Teller approximation with the WKB method. 
The scalar-field mass and the effective potential both affect the QNM spectrum, which may help distinguish rotating black-bounce geometries from Kerr black holes and from more conventional regular black holes through their ringdown signals.

\section{Greybody factors}\label{GBFs}

The greybody factor provides complementary information to the QNM spectrum.
While QNMs describe the free damped oscillations of the system, the greybody
factor characterizes how an incident wave is filtered by the effective
potential before being absorbed by the black hole \cite{Page:1976ki}. Therefore, it is useful to
study the scattering properties of the scalar field in the same background.

We use the radial equation derived in Sec.~III, namely Eq.~(19), together with
the tortoise coordinate in Eq.~(18) and the effective potential in Eq.~(20).
Since the effective potential in the rotating case depends explicitly on the
real frequency $\omega$, the scattering problem has to be solved separately for
each incident frequency. In other words, for each value of $\omega$, we first
construct the corresponding potential $V_{\rm eff}(r;\omega)$ and then
numerically integrate the radial scattering equation.

For the black-hole cases, the boundary conditions are chosen as the usual
scattering boundary conditions. Near the event horizon, the physical solution
is purely ingoing,
\begin{equation}
\psi \sim e^{-i(\omega-m\Omega_H)r_*}, \qquad r\rightarrow r_h ,
\label{eq:gbf_horizon_bc}
\end{equation}
where
\begin{equation}
\Omega_H=\frac{a}{r_h^2+a^2+p^2}
\label{eq:gbf_omega_h}
\end{equation}
is the angular velocity of the horizon. At spatial infinity, the solution is
decomposed into an incoming wave and an outgoing wave,
\begin{equation}
\psi \sim A_{\rm in}e^{-ikr_*}+A_{\rm out}e^{ikr_*},
\qquad r\rightarrow \infty ,
\label{eq:gbf_inf_bc}
\end{equation}
where
\begin{equation}
k=\sqrt{\omega^2-\mu^2}.
\label{eq:gbf_k}
\end{equation}
For a massless scalar field, one has $k=\omega$. After normalizing the
transmitted wave at the horizon to unity, the greybody factor is obtained from
the ratio between the transmitted flux at the horizon and the incident flux at
infinity:
\begin{equation}
\Gamma_{lm}(\omega)=
\frac{\omega-m\Omega_H}{k\,|A_{\rm in}|^2}.
\label{eq:gbf_definition}
\end{equation}
In the non-superradiant regime, $\omega>m\Omega_H$, this quantity is positive
and can be interpreted as the absorption probability of the incident scalar
wave. In the superradiant regime, $\omega<m\Omega_H$, it may become negative,
which corresponds to wave amplification.

\begin{figure*}[h]
\centering
\begin{minipage}[t]{0.48\textwidth}
\centering
\includegraphics[width=\textwidth]{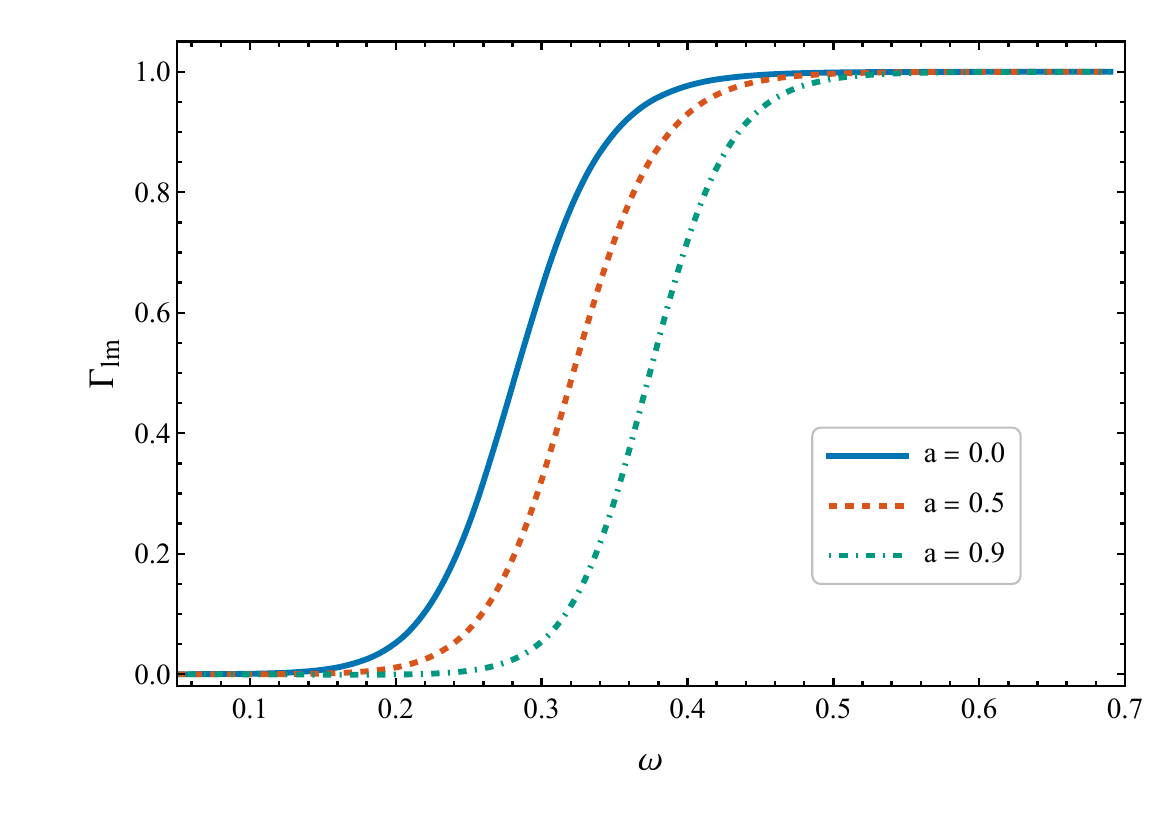}
\par\vspace{0.1mm}
{\small (a)}
\end{minipage}
\hfill
\begin{minipage}[t]{0.48\textwidth}
\centering
\includegraphics[width=\textwidth]{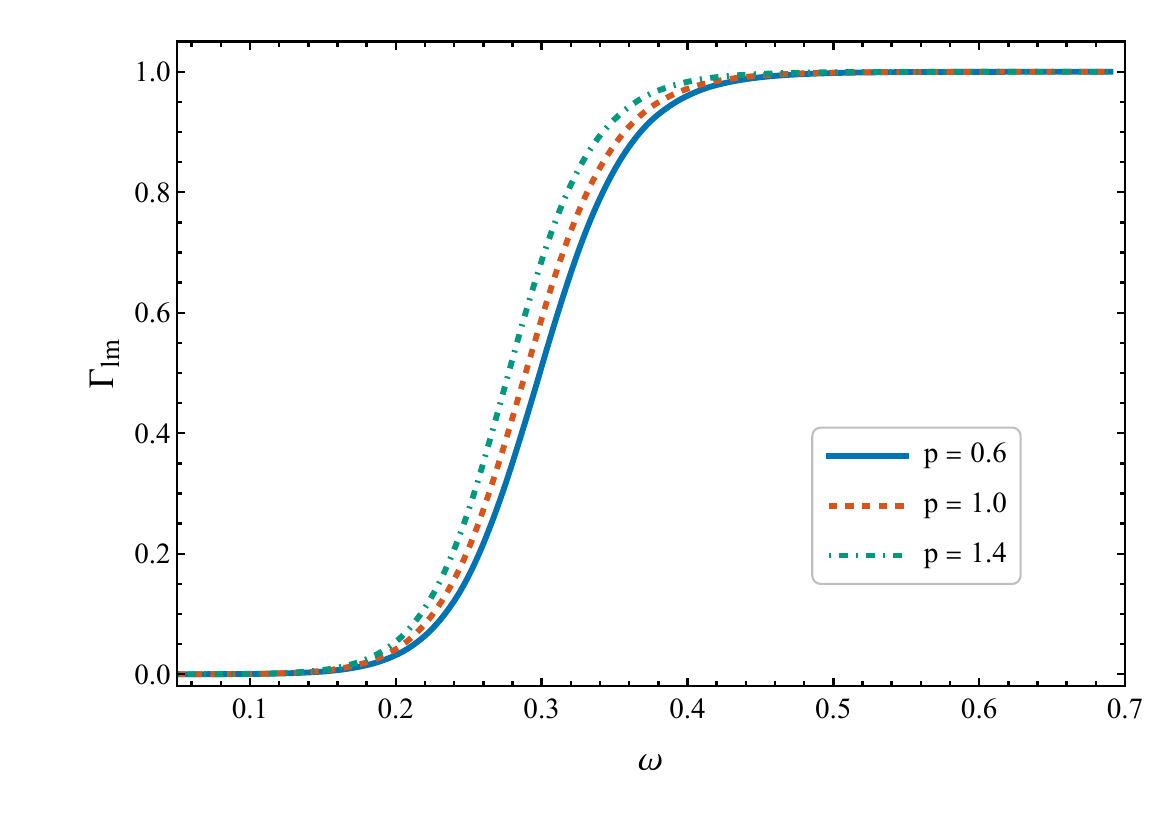}
\par\vspace{0.1mm}
{\small (b)}
\end{minipage}

\caption{
Greybody factors of the Kerr-like black-bounce
spacetime. Panel (a) shows the dependence on the spin parameter $a$, with
$M=1$, $p=1$, $m=1$, $\Lambda_{lm}=2$. Panel (b) shows the
dependence on the bounce parameter $p$, with $M=1$, $a=0.1$, $m=1$,
$\Lambda_{lm}=2$. }
\label{fig:gbf}
\end{figure*}

Fig.~\ref{fig:gbf} shows the greybody factors of the Kerr-like black-bounce spacetime for different values of the spin parameter $a$ and the bounce parameter $p$.
In Fig.~\ref{fig:gbf}(a), we fix $M=1$, $p=1$, $m=1$,
$\Lambda_{lm}=2$, and vary the spin parameter $a$. For all
values of $a$, the greybody factor is strongly suppressed in the
low-frequency region and tends to unity in the high-frequency region. This is
the standard behavior of a wave scattered by a potential barrier:
low-frequency modes have long wavelengths and are mostly reflected, whereas
high-frequency modes can pass through the barrier with a much larger
probability. Increasing $a$ shifts the transition region toward larger
$\omega$. Equivalently, at a fixed intermediate frequency, the greybody
factor decreases as the spin parameter increases. This means that rotation
makes the effective barrier less transparent to low- and intermediate-frequency
scalar waves in the parameter range considered here. Therefore, the spin
parameter affects not only the QNM spectrum but also the absorption channel of
the scalar perturbation.

In Fig.~\ref{fig:gbf}(b), we fix $M=1$, $a=0.1$, $m=1$,
$\Lambda_{lm}=2$, and vary the bounce parameter $p$. Again,
$\Gamma_{lm}$ increases from nearly zero at low frequency to approximately
unity at high frequency. However, the effect of $p$ is opposite to that of
$a$ in the transition region. A larger value of $p$ shifts the curve toward
smaller $\omega$, which implies a larger transmission probability at the same
intermediate frequency. Hence, in the present parameter range, the
black-bounce deformation enhances the scalar absorption probability. This
shows that the bounce parameter can leave a visible imprint not only on the
ringdown spectrum but also on the scattering properties of the spacetime.

The above behavior can be understood from the structure of the effective
potential. The greybody factor is controlled by the barrier between the horizon
and infinity. When the effective barrier is higher or wider, the incoming wave
is more strongly reflected and the greybody factor is smaller. When the
barrier becomes lower or effectively narrower, the transmission probability
increases. Thus, the shifts of the curves in Fig.~\ref{fig:gbf}(a) and
Fig.~\ref{fig:gbf}(b) reflect the different ways in which the spin parameter
and the bounce parameter deform the scalar effective potential. The parameter
$a$ tends to suppress the transmission of low-frequency modes, while the
parameter $p$ enhances the transmission in the intermediate-frequency region.

\section{Summary and outlook}
\label{sec:summary}

In this work we have analyzed the quasinormal modes of a Kerr-like black-bounce spacetime sourced by massive scalar-field perturbations. Our primary goal was to understand how deviations from the Kerr geometry, encoded in the bounce parameter $p$ and in the underlying black-bounce construction, modify the ringdown spectrum and potentially leave observable signatures in gravitational-wave signals.

Starting from the Kerr-like black-bounce metric obtained via the Newman--Janis algorithm \cite{Xu:2021lff}, and specializing to the case $n=2$ and $k=0$, we derived the scalar-field Klein--Gordon equation in this background. By separating variables, we obtained the radial and angular equations and recast the radial sector into a Schr\"odinger-like form in terms of the tortoise coordinate. This allowed us to identify an effective potential $V_{\rm eff}(r)$, which provides the basis for the P\"oschl--Teller and WKB calculations.

To compute the QNM spectrum, we employed two complementary semi-analytic methods: the P\"oschl--Teller potential approximation and the WKB approach (up to sixth order). We verified that, for the parameter ranges considered, the QNM frequencies obtained from these methods are in reasonable agreement, as summarized in Table~\ref{t1} and \ref{t2}. We then systematically explored the dependence of the real and imaginary parts of the frequencies on the spin $a$, the bounce parameter $p$, the separation constant $\Lambda_{lm}$, and the scalar mass $\mu$, presenting our results in Figs.~\ref{w-a}--\ref{w-p}.

Our main physical findings can be summarized as follows:
\begin{itemize}
    \item For fixed $\mu$ and $p$, increasing the spin parameter $a$ increases the real part of the frequency and decreases the damping rate $-\mathrm{Im}(\omega)$ in the parameter range considered here. 
Thus, faster rotation enhances the oscillation frequency and makes the scalar perturbation longer-lived.
    \item The bounce parameter $p$, which controls the deviation from the Kerr solution and the transition between black-hole-like and wormhole-like configurations, affects the spectrum in a different way. A larger $p$ lowers the oscillation frequency and also decreases the damping rate $-\mathrm{Im}(\omega)$. Thus, the black-bounce deformation makes the perturbation decay more slowly, although its effect on the real part is opposite to that of the spin parameter $a$.
    \item The mass $\mu$ of the scalar field significantly affects the QNM spectrum: massive modes develop a richer structure in the $(\mu,\omega)$ plane and contribute to shifting both the real and imaginary parts of the frequencies. Thus, even at the level of test fields, the mass term should not be neglected when modeling ringdowns in such backgrounds.
\end{itemize}

We also studied the greybody factors of the Kerr-like black-bounce spacetime. 
The greybody factor is suppressed at low frequencies and approaches unity at high frequencies. 
Increasing the spin parameter $a$ reduces the transmission probability of low- and intermediate-frequency modes, whereas increasing the bounce parameter $p$ enhances the absorption probability. 
Thus, greybody factors provide a complementary probe of the black-bounce deformation beyond the QNM spectrum.

Taken together, these results indicate that Kerr-like black-bounce spacetimes can mimic Kerr black holes at the level of global properties while still leaving distinct imprints in frequency-domain and scattering observables, particularly through modified QNM spectra and greybody factors. From the observational perspective, current and future gravitational-wave detectors --- including the LIGO/Virgo/KAGRA network and the planned space-based mission LISA \cite{Barausse:2020rsu} --- offer an exciting opportunity to search for such signatures and to constrain deviations from the Kerr paradigm.

Several natural extensions of this work suggest themselves. One important next step is a full time-domain analysis of scalar (and gravitational) perturbations in Kerr-like black-bounce spacetimes, aimed specifically at resolving the echo structure and quantifying its dependence on $(a,p)$ and on the type of perturbation. Another direction is to move beyond test fields and study axial and polar gravitational perturbations, possibly within a gauge-invariant formalism, in order to connect more directly with the gravitational-wave polarizations measured by detectors. It would also be interesting to investigate superradiant scattering and possible instability windows for massive fields in the Kerr-like black-bounce background, as well as to explore the impact of dark-matter environments and plasma effects on the QNM spectrum.

\begin{acknowledgments}
We are very grateful for Prof. A. Zhidenko useful correspondence. This research was funded by the project of Young Scientific and Technical Talents Development of Education Department of Guizhou Province (No.~[2024] 79), the National Natural Science Foundation of China (No.~12505064), Guizhou Provincial Basic Research Program (Natural Science) Youth Guidance Program  (No.~QN [2025] 365), the Guizhou Provincial Basic Research Program (Natural Science) under Grant No.~QN [2025] 310, the Guizhou Provincial Basic Research Program General Project (No.~MS [2026] 068). A. \"O. would like to acknowledge networking support of the COST Action CA21106 - COSMIC WISPers in the Dark Universe: Theory, astrophysics and experiments (CosmicWISPers), the COST Action CA22113 - Fundamental challenges in theoretical physics (THEORY-CHALLENGES), the COST Action CA21136 - Addressing observational tensions in cosmology with systematics and fundamental physics (CosmoVerse), the COST Action CA23130 - Bridging high and low energies in search of quantum gravity (BridgeQG), and the COST Action CA23115 - Relativistic Quantum Information (RQI) funded by COST (European Cooperation in Science and Technology). A. \"O. also thanks to EMU, TUBITAK, ULAKBIM (Turkiye) and SCOAP3 (Switzerland) for their support.

\end{acknowledgments}

\bibliography{reference}

@article{LIGOScientific:2016aoc,
    author = "Abbott, B. P. and others",
    collaboration = "LIGO Scientific, Virgo",
    title = "{Observation of Gravitational Waves from a Binary Black Hole Merger}",
    eprint = "1602.03837",
    archivePrefix = "arXiv",
    primaryClass = "gr-qc",
    reportNumber = "LIGO-P150914",
    doi = "10.1103/PhysRevLett.116.061102",
    journal = "Phys. Rev. Lett.",
    volume = "116",
    number = "6",
    pages = "061102",
    year = "2016"
}

@article{EventHorizonTelescope:2019dse,
    author = "Akiyama, Kazunori and others",
    collaboration = "Event Horizon Telescope",
    title = "{First M87 Event Horizon Telescope Results. I. The Shadow of the Supermassive Black Hole}",
    eprint = "1906.11238",
    archivePrefix = "arXiv",
    primaryClass = "astro-ph.GA",
    doi = "10.3847/2041-8213/ab0ec7",
    journal = "Astrophys. J. Lett.",
    volume = "875",
    pages = "L1",
    year = "2019"
}

@article{Akiyama:2022tyh,
    author = "Akiyama, Kazunori and others",
    title = "{First Sagittarius A* Event Horizon Telescope Results. I. The Shadow of the Supermassive Black Hole in the Center of the Milky Way}",
    doi = "10.3847/2041-8213/ac6674",
    journal = "Astrophys. J. Lett.",
    volume = "930",
    number = "2",
    pages = "L12",
    year = "2022"
}

@article{Barack:2018yly,
    author = "Barack, Leor and others",
    title = "{Black holes, gravitational waves and fundamental physics: a roadmap}",
    eprint = "1806.05195",
    archivePrefix = "arXiv",
    primaryClass = "gr-qc",
    doi = "10.1088/1361-6382/ab0587",
    journal = "Class. Quant. Grav.",
    volume = "36",
    number = "14",
    pages = "143001",
    year = "2019"
}

@article{Simpson:2021zfl,
    author = "Simpson, Alex and Visser, Matt",
    title = "{Astrophysically viable Kerr-like spacetime}",
    eprint = "2112.04647",
    archivePrefix = "arXiv",
    primaryClass = "gr-qc",
    doi = "10.1103/PhysRevD.105.064065",
    journal = "Phys. Rev. D",
    volume = "105",
    number = "6",
    pages = "064065",
    year = "2022"
}

@article{Pantig:2022toh,
    author = {Pantig, Reggie C. and \"Ovg\"un, Ali},
    title = "{Dark matter effect on the weak deflection angle by black holes at the center of Milky Way and M87 galaxies}",
    eprint = "2201.03365",
    archivePrefix = "arXiv",
    primaryClass = "gr-qc",
    doi = "10.1140/epjc/s10052-022-10319-8",
    journal = "Eur. Phys. J. C",
    volume = "82",
    number = "5",
    pages = "391",
    year = "2022"
}

@article{Wang:2001tk,
    author = "Wang, Bin and Abdalla, Elcio and Mann, Robert B.",
    title = "{Scalar wave propagation in topological black hole backgrounds}",
    eprint = "hep-th/0107243",
    archivePrefix = "arXiv",
    doi = "10.1103/PhysRevD.65.084006",
    journal = "Phys. Rev. D",
    volume = "65",
    pages = "084006",
    year = "2002"
}

@article{Kubiznak:2012wp,
    author = "Kubiznak, David and Mann, Robert B.",
    title = "{P-V criticality of charged AdS black holes}",
    eprint = "1205.0559",
    archivePrefix = "arXiv",
    primaryClass = "hep-th",
    doi = "10.1007/JHEP07(2012)033",
    journal = "JHEP",
    volume = "07",
    pages = "033",
    year = "2012"
}

@article{Gunasekaran:2012dq,
    author = "Gunasekaran, Sharmila and Mann, Robert B. and Kubiznak, David",
    title = "{Extended phase space thermodynamics for charged and rotating black holes and Born-Infeld vacuum polarization}",
    eprint = "1208.6251",
    archivePrefix = "arXiv",
    primaryClass = "hep-th",
    reportNumber = "PI-STRONGGRV-291",
    doi = "10.1007/JHEP11(2012)110",
    journal = "JHEP",
    volume = "11",
    pages = "110",
    year = "2012"
}

@article{Kerner:2008qv,
    author = "Kerner, Ryan and Mann, Robert B.",
    title = "{Charged Fermions Tunnelling from Kerr-Newman Black Holes}",
    eprint = "0803.2246",
    archivePrefix = "arXiv",
    primaryClass = "hep-th",
    doi = "10.1016/j.physletb.2008.06.012",
    journal = "Phys. Lett. B",
    volume = "665",
    pages = "277--283",
    year = "2008"
}

@article{Akhmedov:2006pg,
    author = "Akhmedov, Emil T. and Akhmedova, Valeria and Singleton, Douglas",
    title = "{Hawking temperature in the tunneling picture}",
    eprint = "hep-th/0608098",
    archivePrefix = "arXiv",
    reportNumber = "ITEP-TH-38-06",
    doi = "10.1016/j.physletb.2006.09.028",
    journal = "Phys. Lett. B",
    volume = "642",
    pages = "124--128",
    year = "2006"
}

@article{Singleton:2011vh,
    author = "Singleton, Douglas and Wilburn, Steve",
    title = "{Hawking radiation, Unruh radiation and the equivalence principle}",
    eprint = "1102.5564",
    archivePrefix = "arXiv",
    primaryClass = "gr-qc",
    doi = "10.1103/PhysRevLett.107.081102",
    journal = "Phys. Rev. Lett.",
    volume = "107",
    pages = "081102",
    year = "2011"
}

@article{Pantig2022,
title = {Shadow and weak deflection angle of extended uncertainty principle black hole surrounded with dark matter},
journal = {Annals of Physics},
volume = {436},
pages = {168722},
year = {2022},
author = {Reggie C. Pantig and Paul K. Yu and Emmanuel T. Rodulfo and Ali \"Ovg\"un},
}

@article{Barausse:2020rsu,
    author = "Barausse, Enrico and others",
    title = "{Prospects for Fundamental Physics with LISA}",
    eprint = "2001.09793",
    archivePrefix = "arXiv",
    primaryClass = "gr-qc",
    doi = "10.1007/s10714-020-02691-1",
    journal = "Gen. Rel. Grav.",
    volume = "52",
    number = "8",
    pages = "81",
    year = "2020"
}

@article{Simpson:2021dyo,
    author = "Simpson, Alex and Visser, Matt",
    title = "{The eye of the storm: a regular Kerr black hole}",
    eprint = "2111.12329",
    archivePrefix = "arXiv",
    primaryClass = "gr-qc",
    doi = "10.1088/1475-7516/2022/03/011",
    journal = "JCAP",
    volume = "03",
    number = "03",
    pages = "011",
    year = "2022"
}

@article{Martinelli:2021hir,
    author = "Martinelli, Matteo and Casas, Santiago",
    title = "{Cosmological Tests of Gravity: A Future Perspective}",
    eprint = "2112.10675",
    archivePrefix = "arXiv",
    primaryClass = "astro-ph.CO",
    doi = "10.3390/universe7120506",
    journal = "Universe",
    volume = "7",
    number = "12",
    pages = "506",
    year = "2021"
}

@article{Chen:2021cts,
    author = "Chen, Che-Yu and Bouhmadi-L\'opez, Mariam and Chen, Pisin",
    title = "{Lessons from black hole quasinormal modes in modified gravity}",
    eprint = "2103.01249",
    archivePrefix = "arXiv",
    primaryClass = "gr-qc",
    doi = "10.1140/epjp/s13360-021-01227-z",
    journal = "Eur. Phys. J. Plus",
    volume = "136",
    number = "2",
    pages = "253",
    year = "2021"
}

@article{Berti:2015itd,
    author = "Berti, Emanuele and others",
    title = "{Testing General Relativity with Present and Future Astrophysical Observations}",
    eprint = "1501.07274",
    archivePrefix = "arXiv",
    primaryClass = "gr-qc",
    doi = "10.1088/0264-9381/32/24/243001",
    journal = "Class. Quant. Grav.",
    volume = "32",
    pages = "243001",
    year = "2015"
}

@article{Cimdiker:2021cpz,
    author = {\c{C}imdiker, \.Irfan and Demir, Durmu\c{s} and \"Ovg\"un, Ali},
    title = "{Black hole shadow in symmergent gravity}",
    eprint = "2110.11904",
    archivePrefix = "arXiv",
    primaryClass = "gr-qc",
    doi = "10.1016/j.dark.2021.100900",
    journal = "Phys. Dark Univ.",
    volume = "34",
    pages = "100900",
    year = "2021"
}

@article{Cardoso:2016rao,
    author = "Cardoso, Vitor and Franzin, Edgardo and Pani, Paolo",
    title = "{Is the gravitational-wave ringdown a probe of the event horizon?}",
    eprint = "1602.07309",
    archivePrefix = "arXiv",
    primaryClass = "gr-qc",
    doi = "10.1103/PhysRevLett.116.171101",
    journal = "Phys. Rev. Lett.",
    volume = "116",
    number = "17",
    pages = "171101",
    year = "2016",
    note = "[Erratum: Phys.Rev.Lett. 117, 089902 (2016)]"
    
    
}

@article{Cardoso:2016oxy,
    author = "Cardoso, Vitor and Hopper, Seth and Macedo, Caio F. B. and Palenzuela, Carlos and Pani, Paolo",
    title = "{Gravitational-wave signatures of exotic compact objects and of quantum corrections at the horizon scale}",
    eprint = "1608.08637",
    archivePrefix = "arXiv",
    primaryClass = "gr-qc",
    doi = "10.1103/PhysRevD.94.084031",
    journal = "Phys. Rev. D",
    volume = "94",
    number = "8",
    pages = "084031",
    year = "2016"
}

@article{Cardoso:2017cqb,
    author = "Cardoso, Vitor and Pani, Paolo",
    title = "{Tests for the existence of black holes through gravitational wave echoes}",
    eprint = "1709.01525",
    archivePrefix = "arXiv",
    primaryClass = "gr-qc",
    doi = "10.1038/s41550-017-0225-y",
    journal = "Nature Astron.",
    volume = "1",
    number = "9",
    pages = "586--591",
    year = "2017"
}

@article{Maggio:2019zyv,
    author = "Maggio, Elisa and Testa, Adriano and Bhagwat, Swetha and Pani, Paolo",
    title = "{Analytical model for gravitational-wave echoes from spinning remnants}",
    eprint = "1907.03091",
    archivePrefix = "arXiv",
    primaryClass = "gr-qc",
    doi = "10.1103/PhysRevD.100.064056",
    journal = "Phys. Rev. D",
    volume = "100",
    number = "6",
    pages = "064056",
    year = "2019"
}

@article{Qian:2021aju,
    author = "Qian, Wei-Liang and Lin, Kai and Kuang, Xiao-Mei and Wang, Bin and Yue, Rui-Hong",
    title = "{Quasinormal modes in two-photon autocorrelation and the geometric-optics approximation}",
    eprint = "2109.02844",
    archivePrefix = "arXiv",
    primaryClass = "gr-qc",
    doi = "10.1140/epjc/s10052-022-10155-w",
    journal = "Eur. Phys. J. C",
    volume = "82",
    number = "3",
    pages = "188",
    year = "2022"
}

@article{Kuang:2017cgt,
    author = "Kuang, Xiao-Mei and Wu, Jian-Pin",
    title = "{Thermal transport and quasi-normal modes in Gauss\textendash{}Bonnet-axions theory}",
    eprint = "1702.01490",
    archivePrefix = "arXiv",
    primaryClass = "hep-th",
    doi = "10.1016/j.physletb.2017.04.045",
    journal = "Phys. Lett. B",
    volume = "770",
    pages = "117--123",
    year = "2017"
}

@article{Fernando:2016ftj,
    author = "Fernando, Sharmanthie",
    title = "{Quasinormal modes of dilaton-de Sitter black holes: scalar perturbations}",
    eprint = "1601.06407",
    archivePrefix = "arXiv",
    primaryClass = "gr-qc",
    doi = "10.1007/s10714-016-2020-y",
    journal = "Gen. Rel. Grav.",
    volume = "48",
    number = "3",
    pages = "24",
    year = "2016"
}

@article{Fernando:2012yw,
    author = "Fernando, Sharmanthie and Correa, Juan",
    title = "{Quasinormal Modes of Bardeen Black Hole: Scalar Perturbations}",
    eprint = "1208.5442",
    archivePrefix = "arXiv",
    primaryClass = "gr-qc",
    doi = "10.1103/PhysRevD.86.064039",
    journal = "Phys. Rev. D",
    volume = "86",
    pages = "064039",
    year = "2012"
}

@article{Fernando:2003ai,
    author = "Fernando, Sharmanthie",
    title = "{Quasinormal modes of charged dilaton black holes in (2+1)-dimensions}",
    eprint = "hep-th/0306214",
    archivePrefix = "arXiv",
    reportNumber = "NKU-03-SF2",
    doi = "10.1023/B:GERG.0000006694.68399.c9",
    journal = "Gen. Rel. Grav.",
    volume = "36",
    pages = "71--82",
    year = "2004"
}

@article{Fernando:2009tv,
    author = "Fernando, Sharmanthie",
    title = "{Spinning Dilaton Black Holes in 2 +1 Dimensions: Quasi-normal Modes and the Area Spectrum}",
    eprint = "0903.0088",
    archivePrefix = "arXiv",
    primaryClass = "hep-th",
    reportNumber = "NKU-09-SF1",
    doi = "10.1103/PhysRevD.79.124026",
    journal = "Phys. Rev. D",
    volume = "79",
    pages = "124026",
    year = "2009"
}

@article{Fernando:2008hb,
    author = "Fernando, Sharmanthie",
    title = "{Quasinormal modes of charged scalars around dilaton black holes in 2+1 dimensions: Exact frequencies}",
    eprint = "0802.3321",
    archivePrefix = "arXiv",
    primaryClass = "hep-th",
    reportNumber = "NKU-08-SF1",
    doi = "10.1103/PhysRevD.77.124005",
    journal = "Phys. Rev. D",
    volume = "77",
    pages = "124005",
    year = "2008"
}

@article{Fernando:2014gda,
    author = "Fernando, Sharmanthie and Clark, Tyler",
    title = "{Black holes in massive gravity: quasi-normal modes of scalar perturbations}",
    eprint = "1411.6537",
    archivePrefix = "arXiv",
    primaryClass = "gr-qc",
    doi = "10.1007/s10714-014-1834-8",
    journal = "Gen. Rel. Grav.",
    volume = "46",
    number = "12",
    pages = "1834",
    year = "2014"
}

@article{Berti:2009kk,
    author = "Berti, Emanuele and Cardoso, Vitor and Starinets, Andrei O.",
    title = "{Quasinormal modes of black holes and black branes}",
    eprint = "0905.2975",
    archivePrefix = "arXiv",
    primaryClass = "gr-qc",
    doi = "10.1088/0264-9381/26/16/163001",
    journal = "Class. Quant. Grav.",
    volume = "26",
    pages = "163001",
    year = "2009"
}

@article{Cardoso:2008bp,
    author = "Cardoso, Vitor and Miranda, Alex S. and Berti, Emanuele and Witek, Helvi and Zanchin, Vilson T.",
    title = "{Geodesic stability, Lyapunov exponents and quasinormal modes}",
    eprint = "0812.1806",
    archivePrefix = "arXiv",
    primaryClass = "hep-th",
    doi = "10.1103/PhysRevD.79.064016",
    journal = "Phys. Rev. D",
    volume = "79",
    number = "6",
    pages = "064016",
    year = "2009"
}

@article{Andersson:2003fh,
    author = "Andersson, N and Howls, C. J.",
    title = "{The Asymptotic quasinormal mode spectrum of nonrotating black holes}",
    eprint = "gr-qc/0307020",
    archivePrefix = "arXiv",
    doi = "10.1088/0264-9381/21/6/021",
    journal = "Class. Quant. Grav.",
    volume = "21",
    pages = "1623--1642",
    year = "2004"
}

@article{Andersson:1996cm,
    author = "Andersson, Nils",
    title = "{Evolving test fields in a black hole geometry}",
    eprint = "gr-qc/9607064",
    archivePrefix = "arXiv",
    reportNumber = "WUGRAV-96-6, PREPRINT-NO-WUGRAV-96-6",
    doi = "10.1103/PhysRevD.55.468",
    journal = "Phys. Rev. D",
    volume = "55",
    pages = "468--479",
    year = "1997"
}

@article{Andersson:1996xw,
    author = "Andersson, Nils and Onozawa, Hisashi",
    title = "{Quasinormal modes of nearly extreme Reissner-Nordstrom black holes}",
    eprint = "gr-qc/9607054",
    archivePrefix = "arXiv",
    reportNumber = "TIT-HEP-339, COSMO-76, WUGRAV-96-7",
    doi = "10.1103/PhysRevD.54.7470",
    journal = "Phys. Rev. D",
    volume = "54",
    pages = "7470--7475",
    year = "1996"
}

@article{Andersson:1992scr,
    author = "Andersson, Nils and Linn\ae{}us, Staffan",
    title = "{Quasinormal modes of a Schwarzschild black hole: Improved phase-integral treatment}",
    doi = "10.1103/PhysRevD.46.4179",
    journal = "Phys. Rev. D",
    volume = "46",
    number = "10",
    pages = "4179",
    year = "1992"
}

@article{Andersson:1995zk,
    author = "Andersson, N.",
    title = "{Excitation of Schwarzschild black hole quasinormal modes}",
    doi = "10.1103/PhysRevD.51.353",
    journal = "Phys. Rev. D",
    volume = "51",
    pages = "353--363",
    year = "1995"
}

@article{2022zym,
    author = "Konoplya, Roman A. and Zhidenko, Alexander",
    title = "{Can the abyss swallow gravitational waves or why we do not observe echoes?}",
    eprint = "2203.16635",
    archivePrefix = "arXiv",
    primaryClass = "gr-qc",
    month = "3",
    year = "2022"
}

@article{Konoplya:2018yrp,
    author = "Konoplya, R. A. and Stuchl\'\i{}k, Z. and Zhidenko, A.",
    title = "{Echoes of compact objects: new physics near the surface and matter at a distance}",
    eprint = "1810.01295",
    archivePrefix = "arXiv",
    primaryClass = "gr-qc",
    doi = "10.1103/PhysRevD.99.024007",
    journal = "Phys. Rev. D",
    volume = "99",
    number = "2",
    pages = "024007",
    year = "2019"
}

@article{Churilova:2021tgn,
    author = "Churilova, M. S. and Konoplya, R. A. and Stuchlik, Z. and Zhidenko, A.",
    title = "{Wormholes without exotic matter: quasinormal modes, echoes and shadows}",
    eprint = "2107.05977",
    archivePrefix = "arXiv",
    primaryClass = "gr-qc",
    doi = "10.1088/1475-7516/2021/10/010",
    journal = "JCAP",
    volume = "10",
    pages = "010",
    year = "2021"
}

@article{Okyay:2021nnh,
    author = {Okyay, Mert and \"Ovg\"un, Ali},
    title = "{Nonlinear electrodynamics effects on the black hole shadow, deflection angle, quasinormal modes and greybody factors}",
    eprint = "2108.07766",
    archivePrefix = "arXiv",
    primaryClass = "gr-qc",
    doi = "10.1088/1475-7516/2022/01/009",
    journal = "JCAP",
    volume = "01",
    number = "01",
    pages = "009",
    year = "2022"
}

@article{Gonzalez:2021vwp,
    author = "Gonz\'alez, P. A. and Rinc\'on, \'Angel and Saavedra, Joel and V\'asquez, Yerko",
    title = "{Superradiant instability and charged scalar quasinormal modes for (2+1)-dimensional Coulomb-like AdS black holes from nonlinear electrodynamics}",
    eprint = "2107.08611",
    archivePrefix = "arXiv",
    primaryClass = "gr-qc",
    doi = "10.1103/PhysRevD.104.084047",
    journal = "Phys. Rev. D",
    volume = "104",
    number = "8",
    pages = "084047",
    year = "2021"
}

@article{Panotopoulos:2020mii,
    author = "Panotopoulos, Grigoris and Rinc\'on, \'Angel",
    title = "{Quasinormal spectra of scale-dependent Schwarzschild\textendash{}de Sitter black holes}",
    eprint = "2011.02860",
    archivePrefix = "arXiv",
    primaryClass = "gr-qc",
    doi = "10.1016/j.dark.2020.100743",
    journal = "Phys. Dark Univ.",
    volume = "31",
    pages = "100743",
    year = "2021"
}

@article{Panotopoulos:2019gtn,
    author = "Panotopoulos, Grigoris and Rinc\'on, \'Angel",
    title = "{Quasinormal modes of five-dimensional black holes in non-commutative geometry}",
    eprint = "1910.08538",
    archivePrefix = "arXiv",
    primaryClass = "gr-qc",
    doi = "10.1140/epjp/s13360-019-00016-z",
    journal = "Eur. Phys. J. Plus",
    volume = "135",
    number = "1",
    pages = "33",
    year = "2020"
}

@article{Panotopoulos:2019qjk,
    author = "Panotopoulos, Grigoris and Rinc\'on, \'Angel",
    title = "{Quasinormal modes of regular black holes with non linear-Electrodynamical sources}",
    eprint = "1904.10847",
    archivePrefix = "arXiv",
    primaryClass = "gr-qc",
    doi = "10.1140/epjp/i2019-12686-x",
    journal = "Eur. Phys. J. Plus",
    volume = "134",
    number = "6",
    pages = "300",
    year = "2019"
}

@article{Rincon:2018sgd,
    author = "Rinc\'on, \'Angel and Panotopoulos, Grigoris",
    title = "{Quasinormal modes of scale dependent black holes in ( 1+2 )-dimensional Einstein-power-Maxwell theory}",
    eprint = "1801.03248",
    archivePrefix = "arXiv",
    primaryClass = "hep-th",
    doi = "10.1103/PhysRevD.97.024027",
    journal = "Phys. Rev. D",
    volume = "97",
    number = "2",
    pages = "024027",
    year = "2018"
}

@article{Ovgun:2018gwt,
    author = {\"Ovg\"un, Al\"\i{} and Jusufi, Kimet},
    title = "{Quasinormal Modes and Greybody Factors of $f(R)$ gravity minimally coupled to a cloud of strings in $2+1$ Dimensions}",
    eprint = "1801.02555",
    archivePrefix = "arXiv",
    primaryClass = "gr-qc",
    doi = "10.1016/j.aop.2018.05.013",
    journal = "Annals Phys.",
    volume = "395",
    pages = "138--151",
    year = "2018"
}

@article{Daghigh:2008jz,
    author = "Daghigh, Ramin G. and Green, Michael D.",
    title = "{Highly Real, Highly Damped, and Other Asymptotic Quasinormal Modes of Schwarzschild-Anti De Sitter Black Holes}",
    eprint = "0808.1596",
    archivePrefix = "arXiv",
    primaryClass = "gr-qc",
    doi = "10.1088/0264-9381/26/12/125017",
    journal = "Class. Quant. Grav.",
    volume = "26",
    pages = "125017",
    year = "2009"
}

@article{Daghigh:2011ty,
    author = "Daghigh, Ramin G. and Green, Michael D.",
    title = "{Validity of the WKB Approximation in Calculating the Asymptotic Quasinormal Modes of Black Holes}",
    eprint = "1112.5397",
    archivePrefix = "arXiv",
    primaryClass = "gr-qc",
    doi = "10.1103/PhysRevD.85.127501",
    journal = "Phys. Rev. D",
    volume = "85",
    pages = "127501",
    year = "2012"
}

@article{Daghigh:2020mog,
    author = "Daghigh, Ramin G. and Green, Michael D. and Morey, Jodin C. and Kunstatter, Gabor",
    title = "{Scalar Perturbations of a Single-Horizon Regular Black Hole}",
    eprint = "2009.02367",
    archivePrefix = "arXiv",
    primaryClass = "gr-qc",
    doi = "10.1103/PhysRevD.102.104040",
    journal = "Phys. Rev. D",
    volume = "102",
    number = "10",
    pages = "104040",
    year = "2020"
}

@article{Zhidenko:2003wq,
    author = "Zhidenko, A.",
    title = "{Quasinormal modes of Schwarzschild de Sitter black holes}",
    eprint = "gr-qc/0307012",
    archivePrefix = "arXiv",
    doi = "10.1088/0264-9381/21/1/019",
    journal = "Class. Quant. Grav.",
    volume = "21",
    pages = "273--280",
    year = "2004"
}

@article{Zhidenko:2005mv,
    author = "Zhidenko, Alexander",
    title = "{Quasi-normal modes of the scalar hairy black hole}",
    eprint = "gr-qc/0510039",
    archivePrefix = "arXiv",
    doi = "10.1088/0264-9381/23/9/024",
    journal = "Class. Quant. Grav.",
    volume = "23",
    pages = "3155--3164",
    year = "2006"
}

@article{Konoplya:2011qq,
    author = "Konoplya, R. A. and Zhidenko, A.",
    title = "{Quasinormal modes of black holes: From astrophysics to string theory}",
    eprint = "1102.4014",
    archivePrefix = "arXiv",
    primaryClass = "gr-qc",
    doi = "10.1103/RevModPhys.83.793",
    journal = "Rev. Mod. Phys.",
    volume = "83",
    pages = "793--836",
    year = "2011"
}

@article{Chabab:2017knz,
    author = "Chabab, M. and El Moumni, H. and Iraoui, S. and Masmar, K.",
    title = "{Phase Transition of Charged-AdS Black Holes and Quasinormal Modes : a Time Domain Analysis}",
    eprint = "1701.00872",
    archivePrefix = "arXiv",
    primaryClass = "hep-th",
    doi = "10.1007/s10509-017-3175-z",
    journal = "Astrophys. Space Sci.",
    volume = "362",
    number = "10",
    pages = "192",
    year = "2017"
}

@article{Lepe:2004kv,
    author = "Lepe, Samuel and Saavedra, Joel",
    title = "{Quasinormal modes, superradiance and area spectrum for 2+1 acoustic black holes}",
    eprint = "gr-qc/0410074",
    archivePrefix = "arXiv",
    reportNumber = "GACG-04-13",
    doi = "10.1016/j.physletb.2005.05.021",
    journal = "Phys. Lett. B",
    volume = "617",
    pages = "174--181",
    year = "2005"
}

@article{Gonzalez:2017shu,
    author = "Gonz\'alez, P. A. and Papantonopoulos, Eleftherios and Saavedra, Joel and V\'asquez, Yerko",
    title = "{Superradiant Instability of Near Extremal and Extremal Four-Dimensional Charged Hairy Black Hole in anti-de Sitter Spacetime}",
    eprint = "1702.00439",
    archivePrefix = "arXiv",
    primaryClass = "gr-qc",
    doi = "10.1103/PhysRevD.95.064046",
    journal = "Phys. Rev. D",
    volume = "95",
    number = "6",
    pages = "064046",
    year = "2017"
}

@article{Lobo:2020ffi,
    author = "Lobo, Francisco S. N. and Rodrigues, Manuel E. and Silva, Marcos V. de S. and Simpson, Alex and Visser, Matt",
    title = "{Novel black-bounce spacetimes: wormholes, regularity, energy conditions, and causal structure}",
    eprint = "2009.12057",
    archivePrefix = "arXiv",
    primaryClass = "gr-qc",
    doi = "10.1103/PhysRevD.103.084052",
    journal = "Phys. Rev. D",
    volume = "103",
    number = "8",
    pages = "084052",
    year = "2021"
}

@article{Simpson:2018tsi,
    author = "Simpson, Alex and Visser, Matt",
    title = "{Black-bounce to traversable wormhole}",
    eprint = "1812.07114",
    archivePrefix = "arXiv",
    primaryClass = "gr-qc",
    doi = "10.1088/1475-7516/2019/02/042",
    journal = "JCAP",
    volume = "02",
    pages = "042",
    year = "2019"
}

@article{Yang:2021cvh,
    author = "Yang, Yi and Liu, Dong and Xu, Zhaoyi and Xing, Yujia and Wu, Shurui and Long, Zheng-Wen",
    title = "{Echoes of novel black-bounce spacetimes}",
    eprint = "2107.06554",
    archivePrefix = "arXiv",
    primaryClass = "gr-qc",
    doi = "10.1103/PhysRevD.104.104021",
    journal = "Phys. Rev. D",
    volume = "104",
    number = "10",
    pages = "104021",
    year = "2021"
}

@article{Xu:2021lff,
    author = "Xu, Zhaoyi and Tang, Meirong",
    title = "{Rotating spacetime: black-bounces and quantum deformed black hole}",
    eprint = "2109.13813",
    archivePrefix = "arXiv",
    primaryClass = "gr-qc",
    doi = "10.1140/epjc/s10052-021-09635-2",
    journal = "Eur. Phys. J. C",
    volume = "81",
    number = "10",
    pages = "863",
    year = "2021"
}

@article{Mazza:2021rgq,
    author = "Mazza, Jacopo and Franzin, Edgardo and Liberati, Stefano",
    title = "{A novel family of rotating black hole mimickers}",
    eprint = "2102.01105",
    archivePrefix = "arXiv",
    primaryClass = "gr-qc",
    doi = "10.1088/1475-7516/2021/04/082",
    journal = "JCAP",
    volume = "04",
    pages = "082",
    year = "2021"
}

@article{Franzin:2022iai,
    author = "Franzin, Edgardo and Liberati, Stefano and Mazza, Jacopo and Dey, Ramit and Chakraborty, Sumanta",
    title = "{Scalar perturbations around rotating regular black holes and wormholes: quasi-normal modes, ergoregion instability and superradiance}",
    eprint = "2201.01650",
    archivePrefix = "arXiv",
    primaryClass = "gr-qc",
    month = "1",
    year = "2022"
   }

@article{Franzin:2021vnj,
    author = "Franzin, Edgardo and Liberati, Stefano and Mazza, Jacopo and Simpson, Alex and Visser, Matt",
    title = "{Charged black-bounce spacetimes}",
    eprint = "2104.11376",
    archivePrefix = "arXiv",
    primaryClass = "gr-qc",
    doi = "10.1088/1475-7516/2021/07/036",
    journal = "JCAP",
    volume = "07",
    pages = "036",
    year = "2021"
}

@article{Ovgun:2020yuv,
    author = {\"Ovg\"un, Ali},
    title = "{Weak Deflection Angle of Black-bounce Traversable Wormholes Using Gauss-Bonnet Theorem in the Dark Matter Medium}",
    eprint = "2011.04423",
    archivePrefix = "arXiv",
    primaryClass = "gr-qc",
    doi = "10.20944/preprints202008.0512.v1",
    journal = "Turk. J. Phys.",
    volume = "44",
    number = "5",
    pages = "465--471",
    year = "2020"
}

@article{Churilova:2019cyt,
    author = "Churilova, M. S. and Stuchlik, Z.",
    title = "{Ringing of the regular black-hole/wormhole transition}",
    eprint = "1911.11823",
    archivePrefix = "arXiv",
    primaryClass = "gr-qc",
    doi = "10.1088/1361-6382/ab7717",
    journal = "Class. Quant. Grav.",
    volume = "37",
    number = "7",
    pages = "075014",
    year = "2020"
}

@article{Bronnikov:2019sbx,
    author = "Bronnikov, Kirill A. and Konoplya, Roman A.",
    title = "{Echoes in brane worlds: ringing at a black hole--wormhole transition}",
    eprint = "1912.05315",
    archivePrefix = "arXiv",
    primaryClass = "gr-qc",
    doi = "10.1103/PhysRevD.101.064004",
    journal = "Phys. Rev. D",
    volume = "101",
    number = "6",
    pages = "064004",
    year = "2020"
}

@article{Lobo:2020kxn,
    author = "Lobo, Francisco S. N. and Simpson, Alex and Visser, Matt",
    title = "{Dynamic thin-shell black-bounce traversable wormholes}",
    eprint = "2003.09419",
    archivePrefix = "arXiv",
    primaryClass = "gr-qc",
    doi = "10.1103/PhysRevD.101.124035",
    journal = "Phys. Rev. D",
    volume = "101",
    number = "12",
    pages = "124035",
    year = "2020"
}

@article{Lima:2020auu,
    author = "Lima, Haroldo C. D. and Benone, Carolina L. and Crispino, Lu\'\i{}s C. B.",
    title = "{Scalar absorption: Black holes versus wormholes}",
    eprint = "2006.03967",
    archivePrefix = "arXiv",
    primaryClass = "gr-qc",
    doi = "10.1103/PhysRevD.101.124009",
    journal = "Phys. Rev. D",
    volume = "101",
    number = "12",
    pages = "124009",
    year = "2020"
}

@article{Nascimento:2020ime,
    author = "Nascimento, J. R. and Petrov, A. Yu. and Porfirio, P. J. and Soares, A. R.",
    title = "{Gravitational lensing in black-bounce spacetimes}",
    eprint = "2005.13096",
    archivePrefix = "arXiv",
    primaryClass = "gr-qc",
    doi = "10.1103/PhysRevD.102.044021",
    journal = "Phys. Rev. D",
    volume = "102",
    number = "4",
    pages = "044021",
    year = "2020"
}

@article{Tsukamoto:2022vkt,
    author = "Tsukamoto, Naoki",
    title = "{Retrolensing by two photon spheres of a black-bounce spacetime}",
    eprint = "2202.09641",
    archivePrefix = "arXiv",
    primaryClass = "gr-qc",
    doi = "10.1103/PhysRevD.105.084036",
    journal = "Phys. Rev. D",
    volume = "105",
    number = "8",
    pages = "084036",
    year = "2022"
}

@article{Tsukamoto:2020bjm,
    author = "Tsukamoto, Naoki",
    title = "{Gravitational lensing in the Simpson-Visser black-bounce spacetime in a strong deflection limit}",
    eprint = "2011.03932",
    archivePrefix = "arXiv",
    primaryClass = "gr-qc",
    doi = "10.1103/PhysRevD.103.024033",
    journal = "Phys. Rev. D",
    volume = "103",
    number = "2",
    pages = "024033",
    year = "2021"
}

@article{Guerrero:2021ues,
    author = "Guerrero, Merce and Olmo, Gonzalo J. and Rubiera-Garcia, Diego and G\'omez, Diego S\'aez-Chill\'on",
    title = "{Shadows and optical appearance of black bounces illuminated by a thin accretion disk}",
    eprint = "2105.15073",
    archivePrefix = "arXiv",
    primaryClass = "gr-qc",
    doi = "10.1088/1475-7516/2021/08/036",
    journal = "JCAP",
    volume = "08",
    pages = "036",
    year = "2021"
}

@article{Ou:2021efv,
    author = "Ou, Min-Yan and Lai, Meng-Yun and Huang, Hyat",
    title = "{Echoes from Asymmetric Wormholes and Black Bounce}",
    eprint = "2111.13890",
    archivePrefix = "arXiv",
    primaryClass = "gr-qc",
    month = "11",
    year = "2021"
}

@article{Bronnikov:2021liv,
    author = "Bronnikov, Kirill A. and Konoplya, Roman A. and Pappas, Thomas D.",
    title = "{General parametrization of wormhole spacetimes and its application to shadows and quasinormal modes}",
    eprint = "2102.10679",
    archivePrefix = "arXiv",
    primaryClass = "gr-qc",
    doi = "10.1103/PhysRevD.103.124062",
    journal = "Phys. Rev. D",
    volume = "103",
    number = "12",
    pages = "124062",
    year = "2021"
}

@article{BLOME1984231,
title = {Quasi-normal oscillations of a schwarzschild black hole},
journal = {Physics Letters A},
volume = {100},
number = {5},
pages = {231-234},
year = {1984},
issn = {0375-9601},
doi = {https://doi.org/10.1016/0375-9601(84)90769-2},
url = {https://www.sciencedirect.com/science/article/pii/0375960184907692},
author = {Hans-Joachim Blome and Bahram Mashhoon},
}

@article{Ferrari:1984zz,
    author = "Ferrari, Valeria and Mashhoon, Bahram",
    title = "{New approach to the quasinormal modes of a black hole}",
    doi = "10.1103/PhysRevD.30.295",
    journal = "Phys. Rev. D",
    volume = "30",
    pages = "295--304",
    year = "1984"
}

@article{Konoplya:2019hlu,
    author = "Konoplya, R. A. and Zhidenko, A. and Zinhailo, A. F.",
    title = "{Higher order WKB formula for quasinormal modes and grey-body factors: recipes for quick and accurate calculations}",
    eprint = "1904.10333",
    archivePrefix = "arXiv",
    primaryClass = "gr-qc",
    doi = "10.1088/1361-6382/ab2e25",
    journal = "Class. Quant. Grav.",
    volume = "36",
    pages = "155002",
    year = "2019"
}

@article{Shankaranarayanan:2022wbx,
    author = "Shankaranarayanan, S. and Johnson, Joseph P.",
    title = "{Modified theories of gravity: Why, how and what?}",
    eprint = "2204.06533",
    archivePrefix = "arXiv",
    primaryClass = "gr-qc",
    doi = "10.1007/s10714-022-02927-2",
    journal = "Gen. Rel. Grav.",
    volume = "54",
    number = "5",
    pages = "44",
    year = "2022"
}

@article{Odintsov:2022cbm,
    author = "Odintsov, Sergei D. and Oikonomou, Vasilis K. and Myrzakulov, Ratbay",
    title = "{Spectrum of Primordial Gravitational Waves in Modified Gravities: A Short Overview}",
    eprint = "2204.00876",
    archivePrefix = "arXiv",
    primaryClass = "gr-qc",
    doi = "10.3390/sym14040729",
    journal = "Symmetry",
    volume = "14",
    number = "4",
    pages = "729",
    year = "2022"
}

@article{Baker:2019gxo,
    author = "Baker, Tessa and others",
    title = "{Novel Probes Project: Tests of gravity on astrophysical scales}",
    eprint = "1908.03430",
    archivePrefix = "arXiv",
    primaryClass = "astro-ph.CO",
    doi = "10.1103/RevModPhys.93.015003",
    journal = "Rev. Mod. Phys.",
    volume = "93",
    number = "1",
    pages = "015003",
    year = "2021"
}

@article{Ferreira:2019xrr,
    author = "Ferreira, Pedro G.",
    title = "{Cosmological Tests of Gravity}",
    eprint = "1902.10503",
    archivePrefix = "arXiv",
    primaryClass = "astro-ph.CO",
    doi = "10.1146/annurev-astro-091918-104423",
    journal = "Ann. Rev. Astron. Astrophys.",
    volume = "57",
    pages = "335--374",
    year = "2019"
}

@article{Iyer:1986np,
    author = "Iyer, Sai and Will, Clifford M.",
    title = "{Black Hole Normal Modes: A {WKB} Approach. 1. Foundations and Application of a Higher Order {WKB} Analysis of Potential Barrier Scattering}",
    reportNumber = "Print-86-1482 (WASH. U., ST. LOUIS)",
    doi = "10.1103/PhysRevD.35.3621",
    journal = "Phys. Rev. D",
    volume = "35",
    pages = "3621",
    year = "1987"
}

@article{Page:1976ki,
    author = "Page, Don N.",
    title = "{Particle Emission Rates from a Black Hole. 2. Massless Particles from a Rotating Hole}",
    doi = "10.1103/PhysRevD.14.3260",
    journal = "Phys. Rev. D",
    volume = "14",
    pages = "3260--3273",
    year = "1976"
}

@article{Schutz:1985km,
    author = "Schutz, Bernard F. and Will, Clifford M.",
    title = "{BLACK HOLE NORMAL MODES: A SEMIANALYTIC APPROACH}",
    reportNumber = "PRINT-85-0063 (WASH.U.,ST.LOUIS)",
    doi = "10.1086/184453",
    journal = "Astrophys. J. Lett.",
    volume = "291",
    pages = "L33--L36",
    year = "1985"
}

@article{Konoplya:2004wg,
    author = "Konoplya, R. A. and Zhidenko, A. V.",
    title = "{Decay of massive scalar field in a Schwarzschild background}",
    eprint = "gr-qc/0411059",
    archivePrefix = "arXiv",
    doi = "10.1016/j.physletb.2005.01.078",
    journal = "Phys. Lett. B",
    volume = "609",
    pages = "377--384",
    year = "2005"
}

@article{Flachi:2012nv,
    author = "Flachi, Antonino and Lemos, Jos{\'e} P. S.",
    title = "{Quasinormal modes of regular black holes}",
    eprint = "1211.6212",
    archivePrefix = "arXiv",
    primaryClass = "gr-qc",
    doi = "10.1103/PhysRevD.87.024034",
    journal = "Phys. Rev. D",
    volume = "87",
    number = "2",
    pages = "024034",
    year = "2013"
}

@article{Toshmatov:2015wga,
    author = "Toshmatov, Bobir and Abdujabbarov, Ahmadjon and Stuchl{\'\i}k, Zden{\v{e}}k and Ahmedov, Bobomurat",
    title = "{Quasinormal modes of test fields around regular black holes}",
    eprint = "1503.05737",
    archivePrefix = "arXiv",
    primaryClass = "gr-qc",
    doi = "10.1103/PhysRevD.91.083008",
    journal = "Phys. Rev. D",
    volume = "91",
    number = "8",
    pages = "083008",
    year = "2015"
}

@article{Khoo:2025qjc,
    author = "Khoo, Fech Scen",
    title = "{Scalar quasinormal modes of rotating regular black holes}",
    eprint = "2503.09390",
    archivePrefix = "arXiv",
    primaryClass = "gr-qc",
    doi = "10.1103/f35l-m8n5",
    journal = "Phys. Rev. D",
    volume = "111",
    number = "12",
    pages = "124025",
    year = "2025"
}

@article{Skvortsova:2024eqi,
    author = "Skvortsova, Milena",
    title = "{Long-lived quasinormal modes of regular and extreme black holes}",
    eprint = "2503.03650",
    archivePrefix = "arXiv",
    primaryClass = "gr-qc",
    doi = "10.1209/0295-5075/adaee2",
    journal = "EPL",
    volume = "149",
    number = "5",
    pages = "59001",
    year = "2025"
}

@article{Peng:2025htj,
    author = "Peng, Yating and Huang, Jia-Hui",
    title = "{Scalar quasinormal modes, Lyapunov exponents and radii of null geodesics of rotating regular black holes}",
    eprint = "2504.21460",
    archivePrefix = "arXiv",
    primaryClass = "gr-qc",
    doi = "10.1140/epjc/s10052-025-14999-w",
    journal = "Eur. Phys. J. C",
    volume = "85",
    number = "11",
    pages = "1312",
    year = "2025"
}

@article{Jha:2023wzo,
    author = "Jha, Sohan Kumar",
    title = "{Photonsphere, shadow, quasinormal modes, and greybody bounds of non-rotating Simpson{\textendash}Visser black hole}",
    eprint = "2309.06454",
    archivePrefix = "arXiv",
    primaryClass = "gr-qc",
    doi = "10.1140/epjp/s13360-023-04384-5",
    journal = "Eur. Phys. J. Plus",
    volume = "138",
    number = "8",
    pages = "757",
    year = "2023"
}

@article{Ahmed:2026ywu,
    author = "Ahmed, Faizuddin and Al-Badawi, Ahmad and Silva, Edilberto O.",
    title = "{Screened Simpson-Visser Black Holes with Asymptotically de-Sitter Core}",
    eprint = "2603.11921",
    archivePrefix = "arXiv",
    primaryClass = "gr-qc",
    month = "3",
    year = "2026"
}

@article{Al-Badawi:2026dof,
    author = "Al-Badawi, Ahmad",
    title = "{Thermodynamics and shadow of Simpson-Visser black hole with phantom global monopoles}",
    eprint = "2602.03888",
    archivePrefix = "arXiv",
    primaryClass = "gr-qc",
    doi = "10.1142/S0219887826501495",
    month = "2",
    year = "2026"
}

@article{Ahmed:2026bwm,
    author = "Ahmed, Faizuddin and Al-Badawi, Ahmad and Fathi, Mohsen",
    title = "{Charged Simpson-Visser AdS Black Holes: Geodesic Structure and Thermodynamic Properties}",
    eprint = "2601.10469",
    archivePrefix = "arXiv",
    primaryClass = "gr-qc",
    month = "1",
    year = "2026"
}

@article{Kumar:2025nio,
    author = "Kumar, Neeraj and Srivastav, Ankur and Channuie, Phongpichit",
    title = "{Simpson-visser-ads black holes: thermodynamics and binary merger}",
    eprint = "2511.21424",
    archivePrefix = "arXiv",
    primaryClass = "gr-qc",
    doi = "10.1140/epjc/s10052-025-15273-9",
    journal = "Eur. Phys. J. C",
    volume = "86",
    number = "1",
    pages = "77",
    year = "2026"
}

@article{Hamil:2025pte,
    author = {Hamil, B. and Al-Badawi, Ahmad and L{\"u}tf{\"u}o{\u{g}}lu, B. C.},
    title = "{Geodesics and scalar perturbations of Schwarzschild black holes embedded in a Dehnen-type dark matter halo with quintessence}",
    eprint = "2505.18611",
    archivePrefix = "arXiv",
    primaryClass = "gr-qc",
    doi = "10.1088/1402-4896/ae0ed7",
    journal = "Phys. Scripta",
    volume = "100",
    number = "10",
    pages = "105008",
    year = "2025"
}

@article{Al-Badawi:2024qpt,
    author = "Al-Badawi, Ahmad and Shaymatov, Sanjar",
    title = "{Quasinormal modes and shadow of Schwarzschild black holes embedded in a Dehnen-type dark matter halo exhibiting a cloud of strings}",
    eprint = "2412.20037",
    archivePrefix = "arXiv",
    primaryClass = "gr-qc",
    doi = "10.1088/1572-9494/ad89b2",
    journal = "Commun. Theor. Phys.",
    volume = "77",
    number = "3",
    pages = "035402",
    year = "2025"
}

@article{Al-Badawi:2024iog,
    author = "Al-Badawi, Ahmad",
    title = "{Quazinormal modes and greybody factor of black hole surrounded by a quintessence in the S-V-T modified gravity as well as shadow}",
    eprint = "2405.12074",
    archivePrefix = "arXiv",
    primaryClass = "gr-qc",
    doi = "10.1088/1402-4896/ad4069",
    journal = "Phys. Scripta",
    volume = "99",
    number = "6",
    pages = "065002",
    year = "2024"
}

@article{Al-Badawi:2024kdw,
    author = "Al-Badawi, Ahmad and Jha, Sohan Kumar and Rahaman, Anisur",
    title = "{Hairy black hole, Fermionic greybody factors, Quasinormal modes, Hawking radiation, Power spectrum and sparsity}",
    eprint = "2402.06739",
    archivePrefix = "arXiv",
    primaryClass = "gr-qc",
    reportNumber = "HEP/123-qed",
    journal = "Eur. Phys. J. C",
    volume = "84",
    pages = "145",
    year = "2024"
}

@article{Barman:2024hwd,
    author = "Barman, Himangshu and Al-Badawi, Ahmad and Jha, Sohan Kumar and Rahaman, Anisur",
    title = "{The quantum corrected Schwarzschild black hole with a linear-quadratic GUP: a comprehensive evaluation}",
    eprint = "2401.14833",
    archivePrefix = "arXiv",
    primaryClass = "gr-qc",
    doi = "10.1088/1475-7516/2024/05/019",
    journal = "JCAP",
    volume = "05",
    pages = "019",
    year = "2024"
}
\bibliographystyle{apsrev4-1}

\end{document}